\documentclass{aa}

\newcommand\lxu{erg s$^{-1}$}
\newcommand\fxu{erg s$^{-1}$ cm$^{-2}$}

\usepackage{graphicx}
\usepackage[hidelinks]{hyperref}
\usepackage{multirow}
\usepackage{tabularx}
\usepackage{longtable,tabu}
\usepackage{lscape}
\usepackage{subfigure}
\usepackage{natbib}
\usepackage{placeins}
\usepackage[dvipsnames]{xcolor}

\bibpunct{(}{)}{;}{a}{}{,} 
\def\logR{\ensuremath{\log R^{\prime}_{\mathrm{HK}}}}

\newcommand{\wsp}{WASP-80}
\newcommand{\xmm}{{\em XMM-Newton}}
\newcommand{\chandra}{{\em Chandra}}
\begin{document}

\title{On the origin of the non-detection of metastable He{\sc i} in the upper atmosphere of the hot Jupiter WASP-80b}

\titlerunning{On the origin of the non-detection of metastable He{\sc i} in WASP-80b}

\authorrunning{Fossati et al.}

\author{L. Fossati\inst{1} \and
        I. Pillitteri\inst{2} \and
        I. F. Shaikhislamov\inst{3} \and
        A. Bonfanti\inst{1} \and
        F. Borsa\inst{4} \and
        I. Carleo\inst{5,6} \and
        G. Guilluy\inst{7} \and
        M. S. Rumenskikh\inst{3,8} }

\institute{Space Research Institute, Austrian Academy of Sciences, Schmiedlstrasse 6, 8042 Graz, Austria\\
\email{Luca.Fossati@oeaw.ac.at}
\and
INAF -- Osservatorio Astronomico di Palermo, P.zza Parlamento 1, I-90134 Palermo, Italy
\and
Institute of Laser Physics, SB RAS, Novosibirsk 630090, Russia
\and
INAF -- Osservatorio Astronomico di Brera, Via E. Bianchi 46, 23807, Merate (LC), Italy
\and
Instituto de Astrof\'{i}sica de Canarias (IAC), 38205 La Laguna, Tenerife, Spain
\and
INAF -- Osservatorio Astronomico di Padova, Vicolo dell'Osservatorio 5, I-35122 Padova, Italy
\and
INAF -- Osservatorio Astrofisico di Torino, Via Osservatorio 20, 10025, Pino Torinese, Italy
\and
Pushkov Institute of Terrestrial Magnetism, Ionosphere and Radiowave Propagation of the Russian Academy of Sciences (IZMIRAN), Troitsk, Moscow 108840, Russia
}

\date{Received date ; Accepted date }

\abstract
{}
{We aim to narrow down the origin of the non-detection of the metastable He{\sc i} triplet at $\approx$10830\,\AA\ obtained for the hot Jupiter WASP-80\,b.}
{We measure the X-ray flux of WASP-80 from archival observations and use it as input to scaling relations accounting for the coronal [Fe/O] abundance ratio to infer the extreme-ultraviolet (EUV) flux in the 200--504\,\AA\ range, which controls the formation of metastable He{\sc i}. We run three dimensional (magneto) hydrodynamic simulations of the expanding planetary upper atmosphere interacting with the stellar wind to study the impact on the He{\sc i} absorption of the stellar high-energy emission, the He/H abundance ratio, the stellar wind, and the possible presence of a planetary magnetic field up to 1\,G.}
{For a low stellar EUV emission, which is favoured by the measured \logR\ value, the He{\sc i} non-detection can be explained by a solar He/H abundance ratio in combination with a strong stellar wind, or by a sub-solar He/H abundance ratio, or by a combination of the two. For a high stellar EUV emission, the non-detection implies a sub-solar He/H abundance ratio. A planetary magnetic field is unlikely to be the cause of the non-detection.}
{The low EUV stellar flux, driven by the low [Fe/O] coronal abundance, is the likely primary cause of the He{\sc i} non-detection. High-quality EUV spectra of nearby stars are urgently needed to improve the accuracy of high-energy emission estimates, which would then enable one to employ the observations to constrain the planetary He/H abundance ratio and the stellar wind strength. This would greatly enhance the information that can be extracted from He{\sc i} atmospheric characterisation observations.}

\maketitle
\section{Introduction}\label{sec:intro}
\citet{seager2000} and \citet{oklopcic2018} showed that the metastable He{\sc i} (2$^3$S) triplet at $\approx$10830\,\AA\ in the near infrared can probe exoplanetary upper atmospheres in alternative to the ultraviolet (UV) band. Thus the He{\sc i} triplet can constrain atmospheric loss that plays a pivotal role in the evolution of exoplanets and in shaping their observed mass-radius distribution \citep[e.g.][]{lopez2013,jin2014,jin2018,owen2017,kubyshkina2018a,modirrousta2020}.

The He{\sc i} atoms lying in the upper atmosphere can be excited to the metastable state either through photoionisation followed by recombination or through collisional excitation from the ground state \citep{andretta1997}. The former mechanism requires that the He{\sc i} atoms are irradiated by high-energy photons at wavelengths shorter than the He{\sc i} ionisation energy ($\sim$504\,\AA\ or 24.6\,eV), while the latter mechanism requires a high density of energetic electrons. \citet{oklopcic2018} showed that in planetary atmospheres the photoionisation and recombination mechanism is significantly more efficient than the collisional excitation mechanism in producing metastable He{\sc i}. Instead, depopulation of the metastable state occurs through ionisation, radiative emission, and electron collisions, where the latter two mechanisms bring a He{\sc i} atom directly to the ground state or to the excited singlet state, which then decays to the ground state.

The relative efficiency of the mechanisms mentioned above, particularly those relying on photoionisation, strongly depend on the shape of the stellar spectral energy distribution (SED) irradiating a planet. \citet{oklopcic2019} showed that planets that are most likely to show metastable He{\sc i} absorption are those in close orbit around active stars with a low near-ultraviolet (NUV; $<$2600\,\AA) emission, that is active K-type stars. This is because an intense stellar X-ray and extreme ultraviolet (EUV; together XUV; $<$912\,\AA) emission enables ionisation of He{\sc i} atoms from the ground state, which can then recombine into the metastable state, while a low stellar NUV emission reduces the photoionisation of metastable He{\sc i} atoms.

Primary transit observations aiming at detecting metastable He{\sc i} absorption have been conducted for about 30 planets, with a positive detection in about ten cases \citep[e.g.][]{spake2018,nortmann2018,salz2018,allart2018,alonso2019,ninan2020}. The observed systems span a wide range of stellar spectral types, from early A- to late M-type, and planetary properties, from small (presumably) rocky planets to gas giants. Some non-detections can be therefore explained by the star not having the appropriate SED \citep[e.g. KELT-9 having a too low XUV flux and too high NUV flux;][]{nortmann2018} and/or by the planet not having a large enough amount of He in its atmosphere \citep[e.g. super-Earths such as Trappist-1b and 55\,Cnc\,e;][]{krishnamurthy2021,zhang2021}. Furthermore, the observations carried out so far have been obtained employing different instruments and techniques, namely ground-based high-resolution spectroscopy, ground- or space-based low-resolution spectrophotometry, and ground-based narrow-band photometry.

However, there have been also unexpected non-detections with the most striking being that of WASP-80b, which is an inflated hot Jupiter orbiting a K-type star. \citet{fossati2022} reported the results of three high-quality transit observations of WASP-80b collected with the GIANO-B high-resolution spectrograph \citep{Oliva2006} obtaining an upper limit on the He{\sc i} absorption of 0.7\% (at the 2$\sigma$ level). This non-detection has been further confirmed by narrow band photometry observations \citep{vissapragada2022}. 

\citet{fossati2022} also presented the results of three-dimensional (3D) hydrodynamic simulations of the upper atmosphere of WASP-80b and of its interaction with the stellar wind. They concluded that stellar wind pressure is unlikely to cause the non-detection and suggested instead that the atmosphere may have a low helium abundance relative to hydrogen (He/H) of at least ten times sub-solar. However, \citet{vissapragada2022} suggested that confinement of the planetary atmosphere by a large-scale magnetic field might be responsible for the non-detection of metastable He{\sc i}.

As mentioned above, one of the key elements controlling the population and depopulation of the metastable 2$^3$S level is the stellar SED and in particular the part of the XUV emission primarily responsible for He{\sc i} photoionisation (i.e. 200--504\,\AA). This part of the stellar SED lies in the EUV band and it is observationally poorly constrained \citep[see e.g.][]{france2019}. \citet{poppenhaeger2022} derived scaling relations enabling one to infer the EUV emission in the 200--504\,\AA\ band on the basis of the stellar X-ray luminosity and [Fe/O] coronal abundance ratio, which could be measured from high enough quality X-ray spectra or inferred from stellar activity and age. \citet{poppenhaeger2022} concluded that for stars of similar X-ray luminosity, young and active stars with [Fe/O]\,$<$\,1, tend to have an EUV emission in the 200--504\,\AA\ band lower than that of old and inactive stars, with [Fe/O]\,$>$\,1.

In this work, we re-analyse the available X-ray spectra of WASP-80, as well as those of other planet-hosts, using the resulting X-ray flux to infer the EUV flux employing the scaling relations of \citet{poppenhaeger2022}. We then use the obtained XUV flux as input to 3D hydrodynamic (HD) simulations to identify the possible origin of the non-detection of metastable He{\sc i} absorption. To test the suggestion of \citet{vissapragada2022}, we also employ 3D magneto hydrodynamic (MHD) modelling to estimate the impact of a planetary magnetic field on metastable He{\sc i} absorption. Finally, we place the results obtained for WASP-80b in the context of metastable He{\sc i} observations carried out for other systems. 

This paper is organised as follows. Section~\ref{sec:wasp80} presents the results of a re-analysis of the \xmm\ X-ray spectra of WASP-80. In Section~\ref{sec:modelling}, we describe the employed modelling scheme, while Section~\ref{sec:results} presents the results of the (M)HD simulations. Section~\ref{sec:discussion} shows a comparison of the results obtained for WASP-80b with those of past detections and non-detections present in the literature. Finally, Section~\ref{sec:conclusions} gathers the conclusions.
\section{High energy emission of WASP-80}\label{sec:wasp80}
%

\wsp\ has been observed twice with \xmm\footnote{\url{https://www.cosmos.esa.int/web/xmm-newton}} for a duration of 17\,ks (obsid 0744940101, P.I. Salz) and 32\,ks (obsid 0764100801, P.I. Wheatley). We retrieved the data from the \xmm\ archive\footnote{\url{http://nxsa.esac.esa.int/nxsa-web/\#search}} and reduced the constituent observation data files (ODFs) with the science analysis software (SAS) version 20.0 to obtain FITS tables of X-ray events calibrated in astrometry, arrival times, energies of events, and quality flags.
 
We selected the events in the 0.3--10\,keV range, with PATTERN $\leq$12 and FLAG\,=\,0 as prescribed by the SAS guide\footnote{\url{https://www.cosmos.esa.int/web/xmm-newton/how-to-use-sas}}. We checked the light curves of events at high energies ($>$10\,keV) to find periods of high background count rate during the observations. Observation 0744940101 was deemed free of high background intervals, while observation 0764100801 was affected by highly variable background, mostly for the pn detector. We retained only 5.3\,ks of the 32\,ks of the pn exposure at the end of this screening. 
\begin{figure}
    \centering
    \resizebox{\columnwidth}{!}{
    \includegraphics{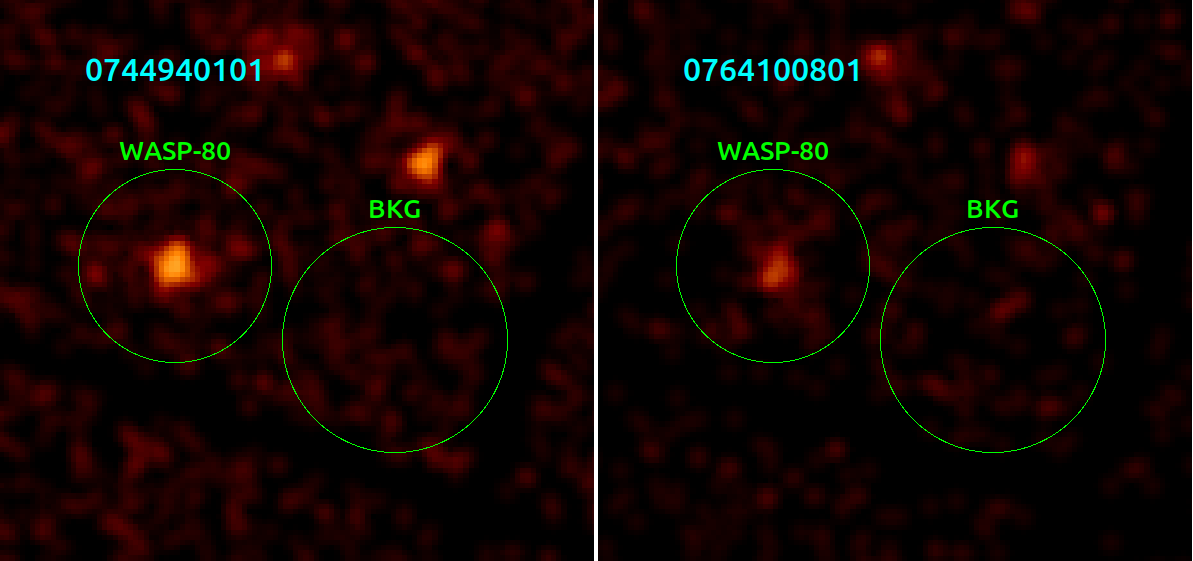}
    }
    \caption{Image of \wsp\ in the two \xmm\ observations (pn detector). The circles show the regions were the spectra of the source and background were accumulated.}
    \label{fig:comparison_pn}
\end{figure}

To accumulate the spectra of WASP-80, we extracted the events related to the target in circular regions of radius 30\arcsec\ centered on the centroid of the X-ray source corresponding to WASP-80. The events used for background subtraction were extracted from a nearby circular region of radius 35\arcsec\ (Figure~\ref{fig:comparison_pn}). The spectra and the related response files were created with SAS. The spectra obtained for each observations from the MOS and pn detectors were combined together to have a summed spectrum with higher count statistics. The response matrices and the background spectra were also combined together with the SAS task {\sc epicspeccombine}\footnote{\url{https://xmm-tools.cosmos.esa.int/external/sas/current/doc/epicspeccombine/index.html}}. The resulting so-called EPIC spectra were then grouped to have a minimum of 30 counts per bin. Summing up all spectra from MOS and pn, the spectra from the first and second exposures had 272 and 370 counts, respectively.

We then used the {\sc xspec}\footnote{\url{https://www.cosmos.esa.int/web/xmm-newton/sas-thread-xspec}} software version 12.11.b to model the two EPIC spectra of both observations and infer N$_{\rm H}$ absorption (i.e. hydrogen column density of the interstellar medium), mean temperatures ($T_1$, $T_2$), emission measure, and flux ($f_X$) in the 0.2--10.0\,keV band. The model was composed by the sum of two absorbed ({\sc tbabs} model) thermal components (APEC). Hydrogen absorption and metal abundances were kept fixed. The resulting best fit parameters are listed in Table~\ref{tab:bestfit}. The choice of this model was motivated by the fact that a simple one-temperature model with an absorbed APEC component gave ambiguous results. In fact, with this model the EPIC spectra could be described either by a low gas absorption of about 5$\times$10$^{20}$\,cm$^{-2}$ and a mean temperature around 0.7\,keV or with a high gas absorption around 10$^{22}$\,cm$^{-2}$ and a cooler temperature $\le0.1$\,keV. As described below, there are solid motivations to discard this second solution at high $N_{\rm H}$.
\begin{table}[t]
\caption{\label{tab:bestfit} Best fit parameters of the EPIC spectra of WASP-80} 
\resizebox{\columnwidth}{!}{
\begin{tabular}{c|cc|c|cc} 
\hline 
\hline 
 & $k_{\rm B}T_1$ & $k_{\rm B}T_2$ & w2/w1 & $f_X$ & $L_X$ \\
 & keV    & keV    &    & log(\fxu) & log(\lxu) \\
\hline
Value & $0.19$ & $0.82$ & $0.8$ & $-13.77$ & $27.7$ \\
C.I. & $0.14 - 0.26$ & $0.71 - 0.95$ & $0.50 - 1.5$ & $-$13.85 to $-$13.69 & 27.6 -- 27.8 \\
\hline
\end{tabular}
}
\tablefoot{Confidence intervals (C.I.) are given at the 90\% level. Unabsorbed flux ($f_X$) and luminosity ($L_X$) are calculated in the 0.2--10\,keV band. The relative weight of the two thermal components is indicated by the ratio between the two APEC components normalization w2/w1 values, being w1 the normalisation of the low temperature component.}
\end{table}

The distance to the star is of about 49.7\,pc \citep{gaiaedr3_2} and at such a distance an interstellar medium hydrogen column density of order $10^{22}$\,cm$^{-2}$ is highly unlikely. To infer a more reliable N$_{\rm H}$ value, we estimated an $E(B-V)$ value of 0.161, to which we arrived by combining the observed $B-V$ color of 1.501\,mag and the estimated intrinsic color $(B-V)_0 = 1.34$\,mag expected for a main sequence star with an effective temperature $T_{\rm eff}$ of 4100\,K such as WASP-80\footnote{\url{https://www.pas.rochester.edu/~emamajek/EEM_dwarf_UBVIJHK_colors_Teff.txt}} \citep{pecaut2013}. We remark that the $(B-V)$ value reported by \citet{salz2015} appears to be too low for a star of spectral type K7 (i.e. $T_{\rm eff}\sim4100$\,K).

From $E(B-V)=0.161$, we inferred a value of A$_V \sim 0.5$\,mag ($R_V=3.1$) and thus $N_{\rm H}\sim10^{20}$\,cm$^{-2}$, which is of the same order of the value obtained from the best fit to the EPIC spectra ($\sim5\times10^{20}$\,cm$^{-2}$). Fixing the hydrogen interstellar absorption to 5$\times$10$^{20}$\,cm$^{-2}$, the best fit gives an unabsorbed flux of $1.7\times10^{-14}$\,\fxu\ in the $0.2-10$\,keV band and an X-ray luminosity of $\sim5\times10^{27}$\,\lxu.

This X-ray luminosity is comparable to those reported by \citet[][$\sim$7$\times$10$^{27}$\,erg\,s$^{-1}$]{salz2015}, \citet[][$\sim$7$\times$10$^{27}$\,erg\,s$^{-1}$]{king2018}, and \citet[][$\sim$5$\times$10$^{27}$\,erg\,s$^{-1}$]{fossati2022}. We remark that the values given by \citet{salz2015} and \citet{king2018} have been obtained considering a larger and less precise distance to the star compared to that measured by GAIA, which was not available at the time. Instead, \citet{fossati2022} rescaled the X-ray luminosity given by \citet{king2018} accounting for the updated stellar distance, but did not consider that a shorter distance implies also a smaller $N_{\rm H}$ value.

We estimated the EUV emission of WASP-80 in the 200--504\,\AA\ wavelength range starting from the measured X-ray flux value and considering the scaling relations of \citet{poppenhaeger2022}, which we recall account for the [Fe/O] abundance in the stellar corona (i.e. high or low relative to solar). The X-ray spectrum of WASP-80 does not allow one to reliably measure the [Fe/O] coronal abundance, therefore we attempted to use the stellar age as proxy. In particular, it is possible to assign a [Fe/O]\,$>$\,1 coronal abundance to inactive stars older than 1\,Gyr and with X-ray luminosity below $10^{28}$ \lxu\ and a [Fe/O]\,$<$\,1 coronal abundance to younger stars. This choice is driven by the low First Ionization Potential (FIP) effect observed in the Sun and in low activity stars \citep{laming2021}, where low FIP elements (such as Fe) appear to be over-abundant with respect to high FIP elements. At the same time, an inverse FIP effect is observed in high activity stars with Fe being under-abundant with respect to low FIP elements. Following \citet{poppenhaeger2022}, this has a strong impact on the presence and strength of emission lines in the EUV band, and thus on the total EUV flux, such that at equal X-ray luminosity young and active stars have an EUV emission in the 200--504\,\AA\ range lower than that of old and inactive stars.

We estimated the age of WASP-80 using the isochrone placement algorithm presented in \citet{bonfanti15,bonfanti16}. This routine interpolates the input parameters (in this case $T_{\mathrm{eff}}$, [Fe/H], and $R_{\star}$) within pre-computed grids of PARSEC\footnote{\textit{PA}dova and T\textit{R}ieste \textit{S}tellar \textit{E}volutionary \textit{C}ode: \url{http://stev.oapd.inaf.it/cgi-bin/cmd}} v1.2S \citep{marigo17} isochrones and tracks to retrieve the best-fit age. For WASP-80, we obtained an age of 1.8$_{-1.8}^{+2.6}$\,Gyr. Therefore, the stellar age is unconstrained, implying that it is not possible to clearly infer the coronal iron abundance, and thus the EUV emission. In the following, we consider that the star can have either a low/high [Fe/O] coronal abundance, and thus a low/high EUV emission (221 and 1520\,erg\,cm$^{-2}$\,s$^{-1}$ at the planetary orbit in the 200--504\,\AA\ wavelength range), and investigate the consequence in terms of formation and possible detection of He{\sc i} metastable absorption in the planetary atmosphere. However, we remark that the measured \logR\ value of about $-$4.04 \citep{fossati2022} implies an age of 12$^{+8}_{-4}$\,Myr \citep{Mamajek2008}, which would therefore suggest that the lower EUV emission value might be preferable. 
\section{Modelling scheme}\label{sec:modelling}
To simulate the upper atmosphere of WASP-80\,b and its interaction with the stellar wind, we employ the 3D (M)HD code described by \citet{ildar2018,khodachenko2021b}. It simulates self-consistently the expansion and escape of the planetary upper atmosphere, which is controlled by the stellar radiative heating and gravitational forces, and its interaction with the surrounding stellar wind, which is also simulated within the model. The extension enabling one to consider the planetary magnetic field strength is achieved adding to the set of hydrodynamic equations the magnetic field induction equation and the Ampere force in the momentum equations for the ionised species. We give here below a brief description of the modelling scheme.

The 3D hydrodynamic multi-fluid numerical model is run in a spherical coordinate system for which the polar axis $Z$ is taken perpendicular to the orbital plane. For presenting the results we also use a Descart frame with the $X$-axis directed along the planet–star line. The simulation reference frame is attached to the planet. Such geometry is suited well to simulate tidally locked planets with the stellar radiation impinging on the planet from just one direction, but we remark that the code enables one to simulate also planets with arbitrary rotation. The code solves numerically the continuity, momentum, and energy equations for separate components, which can be written in the following form \citep{ildar2016}
\begin{equation}
\label{eq:continuity}
\frac{\partial{n_j}}{\partial{t}} + \nabla(V_jn_j) = N_{{\rm XUV},j} + N_{{\rm exh},j}\,,
\end{equation}
\begin{multline}
\label{eq:momentum}
m_j \frac{\partial{V_j}}{\partial{t}} + m(V_j \nabla)V_j = -\frac{1}{n_j}\nabla n_jkT_j - \frac{z_j}{n_e}\nabla n_ekT_e - \\ m_j\nabla U - m_j\sum_jC_{ij}^{\nu}(V_j-V_i)\,,
\end{multline}
and
\begin{multline}
\label{eq:energy}
\frac{\partial{T_j}}{\partial{t}} + (V_j\nabla)T_j + (\gamma -1)T_j\nabla V_j = W_{{\rm XUV},j} - \\ \sum_jC_{ij}^{T}(T_j-T_i)\,,
\end{multline}
respectively. In the above equations, $n_j$ is the density of species $j$, $t$ is time, $V_j$ is the velocity of species $j$, $m_j$ is the mass of a particle of species $j$, $T_j$ is the temperature of species $j$, $n_e$ is the electron density, $T_e$ is the electron temperature, $U$ describes the gravitational interaction (see below), and $W_{{\rm XUV},j}$ is the heating term for the planetary atmosphere (see below). The terms $N_{{\rm XUV},j}$, $N_{{\rm exh},j}$, $C_{ij}^{\nu}$, and $C_{ij}^{T}$ are the photo-ionisation, charge-exchange, and collisional terms listed in Table~1 of \citet{ildar2016}.

The main processes responsible for the transformation between neutral and ionised particles are photoionisation, electron impact ionisation, and dielectronic recombination, which are included in the term $N_{{\rm XUV},j}$ in Equation~(\ref{eq:continuity}) and are applied to all species. Photoionisation also results in heating of the planetary gas through impacts with the produced photoelectrons. The corresponding heating term $W_{{\rm XUV},j}$ in Equation~(\ref{eq:energy}) \citep[see][]{ildar2014,ildar2016,khodachenko2015} comprises terms derived by integrating the stellar XUV spectrum in the 10--912\,\AA\ wavelength range (e.g. Equation~(4) of \citealt{khodachenko2015} and Equation~(5) of \citealt{ildar2016}). The model assumes that the energy released in the form of photoelectrons is rapidly and equally redistributed among all nearby particles with efficiency $\eta_h$\,=\,0.5\,$\div$\,1. This is a commonly used assumption, which we adopted on the basis of qualitative analyses \citep{ildar2014}. The heating term, which includes also energy loss due to excitation and ionisation of hydrogen atoms, in a simplified form can be written as
\begin{multline}
\label{eq:heating_term}
W_{{\rm XUV},j} = (\gamma-1)n_a[\langle(\hslash\nu-E_{\rm ion})\sigma_{\rm XUV}F_{\rm XUV}\rangle - \\ n_e\nu_{Te}(E_{21}\sigma_{12}+E_{\rm ion}\sigma_{\rm ion})]\,,
\end{multline}
where $\nu$ is the frequency of the stellar irradiation, $E_{\rm ion}$ is the ionisation energy, $\sigma_{\rm XUV}$ is the cross-section to the stellar XUV flux $F_{\rm XUV}$, $E_{21}$ is the excitation energy, and $\sigma_{\rm ion}$ and $\sigma_{12}$ are respectively the ionisation and excitation cross sections averaged over a Maxwellian distribution of electrons.

The model further accounts for resonant charge-exchange collisions ($N_{\rm exh}$), which at low energies has a cross-section of $\sigma_{\rm exc}$\,=\,6$\times$10$^{-15}$\,cm$^{-2}$ that is an order of magnitude larger than the elastic collision cross-section. Experimental data on the differential cross-sections can be found, for example, in \citet{lindsay2005}. Since planetary atoms and protons have different thermal pressure profiles and protons feel electron pressure while atoms do not, when they pass close to each other the charge-exchange between them leads to velocity ($\nu$) and temperature ($T$) exchanges. We describe this process with the collision rate $C_{ij}^{\nu,T}$, where the upper index indicates the value being exchanged. For example, in the momentum equation for planetary protons there is the term $C_{H^+H}^{\nu}$\,=\,$n_{\rm H}^{\rm pw}\sigma_{\rm exh}\nu$, where the interaction velocity $\nu$\,$\approx$\,$\sqrt{V_{Ti}^2+V_{Tj}^2+(V_j-V_i)^2}$ depends in general on the thermal and relative velocities of the interacting fluids, in this specific case protons and neutral atoms of the planetary wind. More accurate expressions for the charge-exchange terms present in the continuity, momentum, and energy equations, obtained by averaging the collision operator over the Maxwell distribution \citep[e.g.][]{meier2012}, differ from those used in our model by less than a factor of a few, which is negligible for the purposes of the simulations.

In the model, we considered the following cross-sections: $\sigma_{\rm XUV}$\,=\,6.3$\times$10$^{-18}$($\lambda$/$\lambda_{\rm thr}$)$^3$\,cm$^2$ as the wavelength dependent XUV ionisation cross-section, $\sigma_{\rm ion}$\,=\,4.0$\times$10$^{-16}$$e^{-E_{\rm ion}/T}$T$^{-1}$\,cm$^2$ as the electron impact ionisation cross-section, $\sigma_{\rm rec}$\,=\,6.7$\times$10$^{-21}T^{-3/2}$\,cm$^2$ as the cross-section for recombination with electrons, and $\sigma_{12}$\,=\,3\,$\times$\,$\sigma_{21}e^{-E_{\rm 21}/T}$\,cm$^2$ and $\sigma_{21}$\,=\,7$\times$10$^{-16}T^{-1}$\,cm$^2$ as the hydrogen excitation and de-excitation cross-sections, respectively, where the temperature is scaled in units of the model’s characteristic temperature (i.e. $T^4$\,K), except in the exponents for the expressions of $\sigma_{\rm ion}$ and $\sigma_{12}$, where the temperature is given in erg. 

For the typical parameters of planetary plasmaspheres, Coulomb collisions with protons effectively couple the minor species’ ions. For example, at $T\,<\,10^4$\,K and $n_{\rm H^+}\,>\,10^6$\,cm$^{-3}$ the collisional equalisation time \citep{braginskii1965} for temperature and momentum
\begin{equation}
(C_{{\rm H^+},j}^{\nu})^{-1}\,=\,\tau_{\rm Coul}\,\approx\,\frac{10^6T^2}{n_{\rm H^+}V_{\rm H^+}}\,\frac{M_i}{m_{\rm p}}
\end{equation}
is about 2\,s for protons and about 8\,s for He. This is several orders of magnitude smaller than the typical gas-dynamic time scale of the problem treated here, which is of the order of 10$^4$\,s. 

The strong coupling of charged particles in the planetary wind on the considered typical spatial scale of the problem (i.e. about R$_{rm p}$; $\sim$10$^{10}$\,cm) is further justified by the presence of a chaotic and sporadic magnetic field in the planetary wind, which affects the relative motion of the ions so that they become coupled through the Lorentz force and exchange their momentum on the time scale of the Larmor period. For the same reason, charged particles can be treated as strongly coupled ones in the hot and rarefied stellar wind as well, even in spite of the fact that Coulomb collisions are negligible there. Therefore, there is no need to calculate the dynamics of every charged component of the plasma fluid species, and we assume all of them to have the same temperature and velocity. Instead, the temperature and velocity of each neutral component is calculated individually by solving the corresponding energy and momentum equations. The neutral hydrogen atoms are more or less coupled to the main flow also by elastic collisions. With a typical cross-section of $>$10$^{-16}$\,cm$^2$, the mean-free path at a density of 10$^6$\,cm$^{-3}$ is comparable to the planetary radius. Besides elastic collisions, charge exchange ensures more efficient coupling between hydrogen atoms and protons \citep{ildar2016,khodachenko2017}. Furthermore, the simulation is simplified by assuming that all charged particles have the same velocity, while each neutral fluid has its own particular velocity, including He{\sc i} and He{\sc i}\,(2$^3$s).

The model further accounts for molecular hydrogen and the corresponding ions \citep[H$_2^+$, H$_3^+$; see][]{khodachenko2015,ildar2018}, which allows more accurate treatment of the inner regions of the planetary thermosphere and cooling by the efficient infrared emitter H$_3^+$. The model enables also the inclusion of minor species, which are described as separate fluids by the corresponding momentum and continuity equations. The population of different ionisation states for each element is calculated assuming the specific photoionisation \citep{verner1996} and recombination rates \citep{leTeuff2000,nahar1997}. We remark that we do not consider chemical reactions among the different minor species, while the list of modelled hydrogen reactions can be found in \citet{khodachenko2015}, and it is comparable to that used in other aeronomy models \citep[e.g.][]{garcia2007,koskinen2007}.

To account for the geometry of the problem, we employ a gravitational potential that accounts for rotational effects of the form
\begin{multline}
\label{eq:gravity}
U = -G\frac{M_{\rm p}}{|\overrightarrow{R}|} - G\frac{M_{\rm s}}{|\overrightarrow{R}-\overrightarrow{R_{\rm s}}|} + G\frac{M_{\rm s}\overrightarrow{R}\overrightarrow{R_{\rm s}}}{|\overrightarrow{R_{\rm s}}|^3} - \frac{1}{2}|\overrightarrow{\Omega}\times\overrightarrow{R}|^2\,,
\end{multline}
where the subscript $\rm s$ indicates the star.

The numerical scheme is explicit and uses an up-wind donor cell method for flux calculations. To achieve second order spatial accuracy for differentials, we consider two grids shifted by half a step along each dimension. One grid is reserved for densities, temperatures, and gravity potential, and the other for velocities. For second order temporal accuracy at each time step, the code calculates at first ($n$,$T$) values using the velocity field $V$ and then recomputes $V$ using the new ($n$,$T$) values. The numerical scheme fully conserves flux and total mass, and conserves the Bernoulli constant along the characteristics. For energy, a simple non-conservative equation is used. The energy conservation is checked by global integration and is used to evaluate the accuracy of the simulation. Usually, the energy is balanced within 25\%. We do not use any particular method to capture the shock between planetary and stellar wind. For the problems under consideration, the accurate position of the shock and high front resolution are not crucial. 

The spatial spherical grid uses uniform step for azimuth angle (in this particular study we employ 96 points along a circumference; i.e. $\Delta\phi$\,=\,0.065). The radial grid is exponential with steps varying linearly with radius as $\Delta r$\,=\,$\Delta r_{\rm min} + (\Delta r_{\rm max}-\Delta r_{\rm min})(r-R_{\rm p})/(R_{\rm max}-R_{\rm p})$, where $r$ is the planetocentric radial distance. At the planetary surface $\Delta r$ is as small as $R_{\rm p}$/200. For $\Delta r_{\rm max}$, we employ a value equal to $\Delta\phi\cdot\,R_{\rm max}$. Therefore, in the shock region of about 20\,$R_{\rm p}$ the resolution is about $R_{\rm p}$. For the polar angle, the grid is quadratic, $\Theta\,=\,\Delta\Theta_{\rm min} i+\alpha i^2$, with the smallest step located at the equatorial plane. Usually, $\Delta\Theta = \Delta\phi$ at the equator and $\Delta\Theta = 2\Delta\phi$ along the polar axis. We note that the exponential radial spacing in the spherical coordinate system allows one to keep the same resolution in all three dimensions, if the azimuthal and latitudinal steps are chosen so that $\Delta\phi\approx\Delta\Theta\approx\Delta r/r$. The influence of spatial resolution has been checked by doubling the number of grid points for each dimension obtaining comparable results.

Each simulation is started from an initial static atmosphere and proceeds in the case of WASP-80b for about 500 dimensionless times (corresponding to about 14 orbits) until the overall planetary mass-loss rate reaches 95\% of its asymptotic level, which is judged to be sufficient to assume that the simulation has reached the steady state.

To compute the column densities along stellar rays, we use integration from the star to each cell in the planetary spherical frame, using along the path the density values interpolated from nearby pixels. The calculated column density is used to determine the attenuation of the stellar XUV flux in each spectral bin of 0.1\,nm. To save numerical time, the radiation transfer is calculated usually each forth step of fluid dynamics and chemistry. We assume optically thin approximation for the photons generated by proton recombination to the ground state, and thus the total recombination coefficients are used. 

The chemical reactions are calculated by direct conversion of the matrix $dn_i = R_{ji}(t,r)n_jn_i$ at each time step and at each pixel. This is not efficient numerically, but eliminates convergence problems due to the widely different reaction rates $R_{ji}$. 

Because of the large scale of the considered system, in most of the area surrounding the planet the dynamics of the magnetic field is assumed to be dissipation-less, which in the numerical model is achieved by taking a sufficiently high, though finite, electric conductivity corresponding to a magnetic Reynolds number of about 10$^5$. This value was found empirically to exceed the numerical diffusion in the magnetic field induction equation. The planetary magnetic dipole moment $m$ is directed perpendicularly to the equatorial plane, which is considered to be coplanar to the ecliptic plane. 

In the case of the MHD simulations, we calculate the magnetic field induction equation assuming that the divergence of the magnetic field is zero. At the inner boundary of the computation domain (i.e. at the optical radius of the planet $r$\,=\,$R_{\rm p}$), we fix the flux of the magnetic field by fixing the radial component of the magnetic dipole field (i.e. $B_{\rm r}$\,=\,constant). Instead, the perturbations of the azimuthal (i.e. toroidal; $B_{\rm f}$) and poloidal ($B_{\perp}$) components of the field obey an open boundary condition that is $\partial_{\rm r}(r\cdot\delta B_{\perp})$, where the symbol $\delta$ indicates the perturbation \citep{khodachenko2021b}. The code has already been used to interpret the observations of metastable He{\sc i} absorption for GJ3470\,b \citep{ildar2021}, WASP-107\,b \citep{khodachenko2021a}, and HD189733\,b \citet{rumenskikh2022}, as well as for HD209458\,b in case it hosts a magnetic field \citep{khodachenko2021b}.
\section{Results}\label{sec:results}
~\
\subsection{Non-magnetised planet}\label{sec:nonmagnetic}
%
\begin{figure}
		\centering
		\includegraphics[width=9cm]{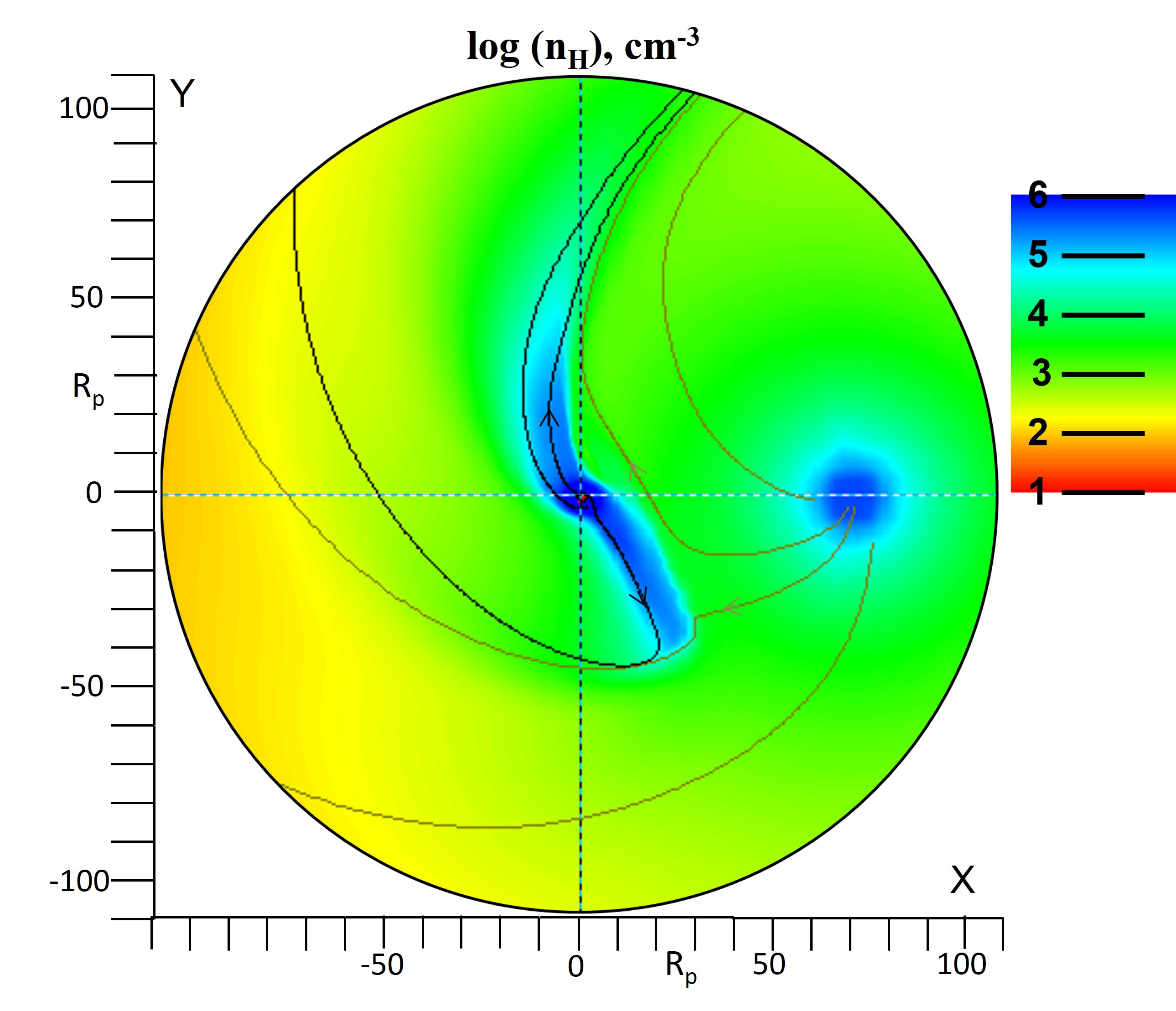}
		\caption{Proton density distribution in the orbital plane of the whole simulated domain for the run computed considering a He/H abundance ratio of 0.01 and an EUV stellar flux at 1\,AU of 0.7\,erg\,cm$^{-2}$\,s$^{-1}$. The planet is at the center of coordinate (0,0) and moves counter-clockwise relative the star, which is located at (76,0). Proton fluid streamlines originated from the planet (black) and from the star (gray) are shown. The axes are scaled in planetary radii.}
		\label{fig:density_map}
\end{figure}
\begin{figure}
		\centering
		\includegraphics[width=9cm]{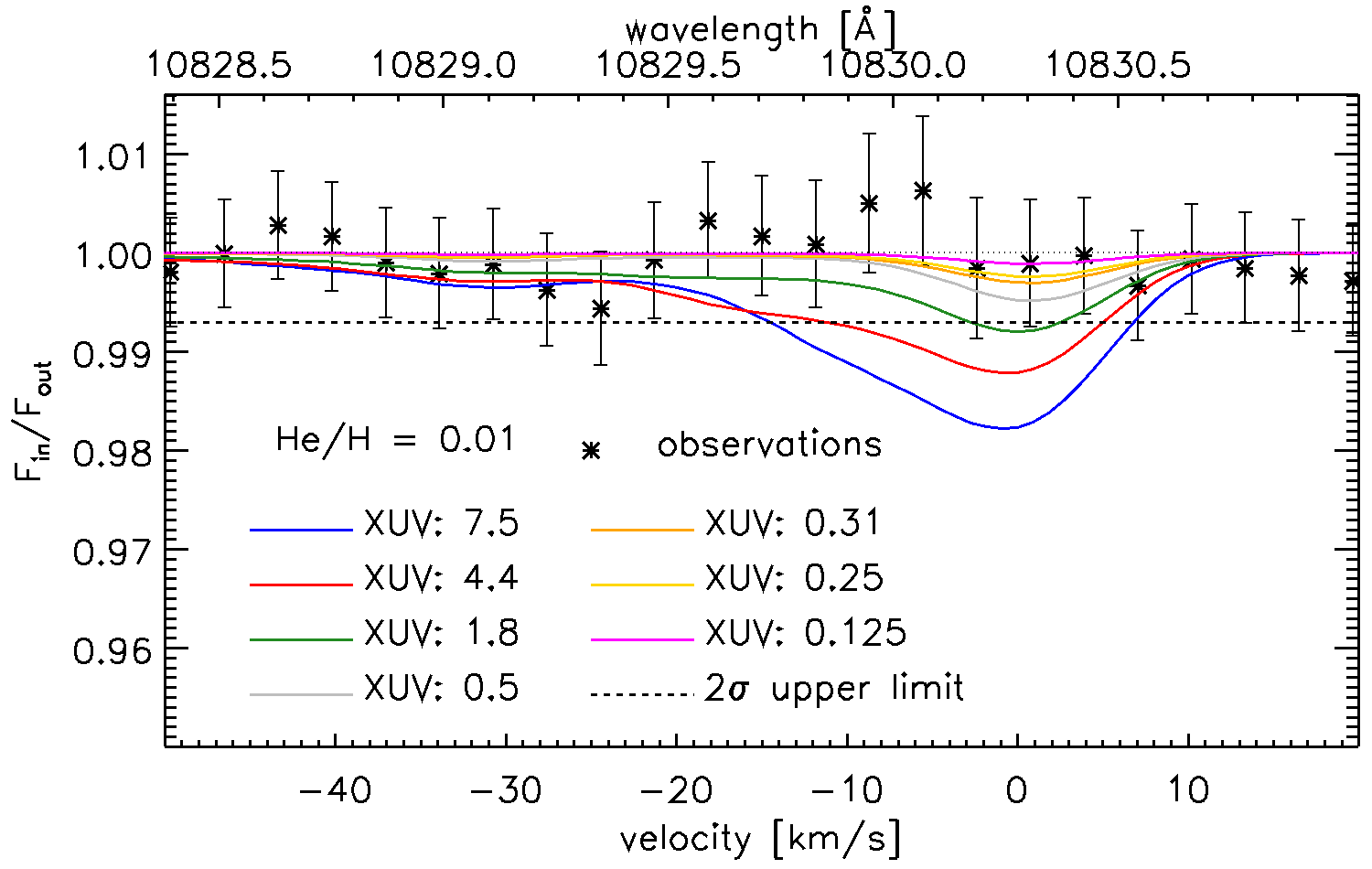}
		\caption{He{\sc i} (2$^3$S) triplet absorption profiles obtained considering a He/H abundance ratio of 0.01 and different values of the stellar XUV flux in comparison with the observations (black asterisks). The XUV flux values at 1\,AU in erg\,cm$^{-2}$\,s$^{-1}$ and the corresponding line colors are given in the legend. The absorption profiles are the result of time averaging from $-$0.1 to $+$0.1 in planetary orbital phase. The zero Doppler-shifted velocity on the $x$ axis corresponds to a wavelength of 10\,830.25\,\AA. The horizontal dashed line marks the 2$\sigma$ upper limit derived from the observations \citep{fossati2022}. The horizontal dotted line at 1.00 is for reference.}
		\label{fig:HeIabsorptionprofiles}
\end{figure}
\begin{table*}[h!]
\renewcommand{\arraystretch}{1.5}
\caption{Input parameters and results of the hydrodynamic simulations.}
\begin{tabular}{c|c c|c|c|c c}
\hline
\hline
He/H & XUV - EUV & XUV - EUV & $\dot M_{\rm sw}$ & $\dot M_{\rm p}$ & Peak & FWHM \\
     & at 1 AU   & at planet &  &  & absorption & absorption \\
     & [erg\,cm$^{-2}$\,s$^{-1}$] & [erg\,cm$^{-2}$\,s$^{-1}$] & [10$^{12}$\,g\,s$^{-1}$] & [10$^{10}$\,g\,s$^{-1}$] & [\%] & [km\,s$^{-1}$] \\
\hline
0.01 & 0.125 - 0.051 &  105.6 -   43.3 & 0.1 & 0.02 & 0.10 & 10 \\
0.01 & 0.25  - 0.10  &  211.3 -   86.6 & 0.1 & 0.06 & 0.24 & 11 \\
0.01 & 0.31  - 0.13  &  262.0 -  107.4 & 0.1 & 0.08 & 0.29 & 11 \\
0.01 & 0.5   - 0.2   &  422.5 -  173.2 & 0.1 & 0.15 & 0.48 & 11 \\
0.01 & 1.8   - 0.7   & 1521.1 -  623.6 & 0.1 & 0.68 & 0.79 & 13 \\
0.01 & 4.4   - 1.8   & 3718.2 - 1524.5 & 0.1 & 1.70 & 1.66 & 20 \\
0.01 & 7.5   - 3.1   & 6337.9 - 2598.5 & 0.1 & 2.80 & 1.81 & 20 \\
\hline
0.03 & 0.31  - 0.13  &  262.0 -  107.4 & 0.1 & 0.07 & 0.72 & 11 \\
0.03 & 1.0   - 0.4   &  845.1 -  712.9 & 0.1 & 0.29 & 1.83 & 12 \\
0.03 & 4.4   - 1.8   & 3718.2 - 1524.5 & 0.1 & 1.58 & 3.23 & 19 \\
\hline
0.1  & 0.125 - 0.051 &  105.6 -   43.3 & 0.1 & 0.01 & 0.44 & 11 \\
0.1  & 0.5   - 0.2   &  422.5 -  173.2 & 0.1 & 0.05 & 1.81 & 12 \\
0.1  & 1.8   - 0.7   & 1521.1 -  623.6 & 0.1 & 0.24 & 4.09 & 13 \\
0.1  & 7.5   - 3.1   & 6337.9 - 2598.5 & 0.1 & 1.72 & 7.79 & 16 \\
\hline
0.1  & 4.4   - 1.8   & 3718.2 - 1524.5 & 20  & 1.47 & 3.23 & 13 \\
0.1  & 0.6   - 0.24  &  507.0 -  202.8 & 20  & 0.37 & 0.95 & 12 \\
\hline
\end{tabular}
\tablefoot{The first column lists the He abundance relative to hydrogen. The second and third columns give the stellar XUV (10--912\,\AA) and EUV (200--504\,\AA) flux at 1\,AU and at the planetary orbit, respectively. The fourth and fifth columns list the stellar wind and planetary mass-loss rates, respectively, while the sixth and seventh columns give the peak and the full width at half maximum (FWHM) of the metastable He{\sc i} absorption (i.e. without the geometric absorption of the planetary disk). The absorption and FWHM values are the result of time averaging from $-$0.1 to $+$0.1 in planetary orbital phase.}
\label{tab:results}
\end{table*}
\begin{figure}[h!]
		\centering
		\includegraphics[width=9cm]{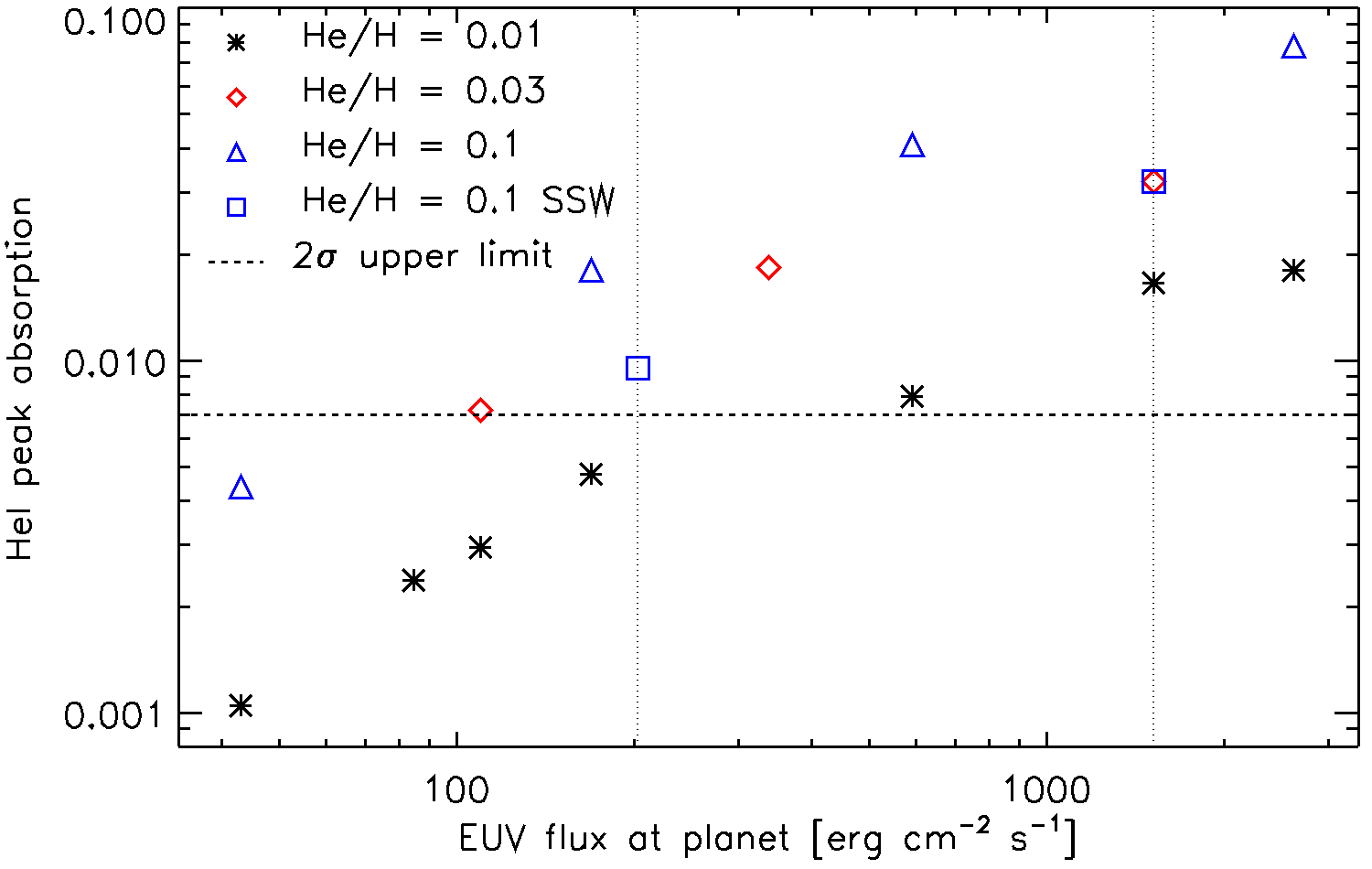}
		\includegraphics[width=9cm]{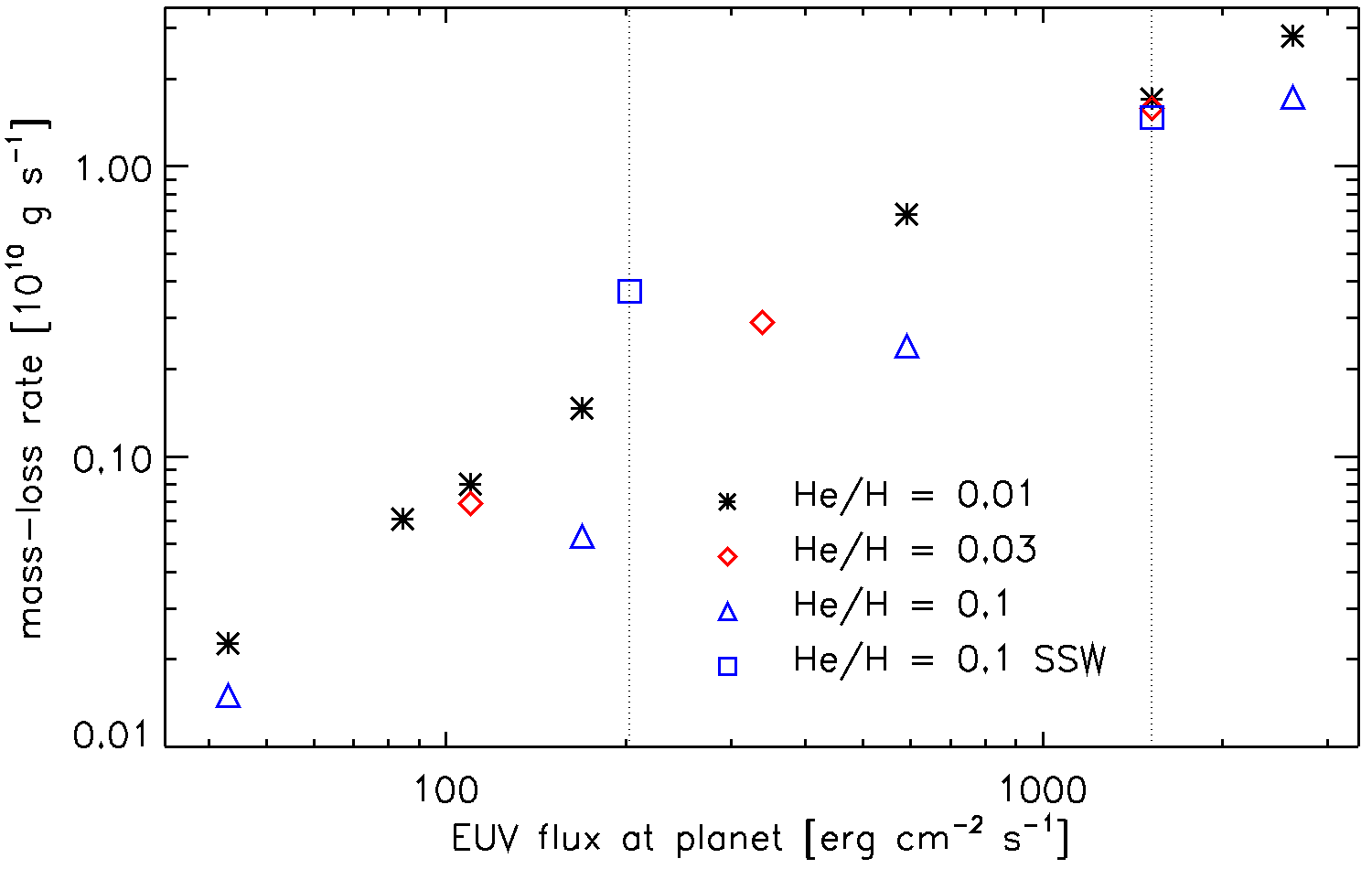}
		\caption{Summary of the results of the hydrodynamic simulations. Top: peak absorption of the He{\sc i} (2$^3$S) triplet resulting from the HD simulations as a function of the stellar EUV emission at the planetary orbital separation, assuming He/H abundance ratios of 0.01 (black asterisks), 0.03 (red rhombs), and 0.1 (blue triangles) and a stellar wind mass-loss rate of 10$^{11}$\,g\,s$^{-1}$. The blue squares are for a He/H abundance ratio of 0.1 and a stronger stellar wind (SSW) of 2$\times$10$^{13}$\,g\,s$^{-1}$. The horizontal dashed line marks the 2$\sigma$ upper limit derived from the observations. The vertical dotted lines correspond to the low and high values of the EUV flux obtained following the re-analysis of the X-ray data and the scaling relations of \citet{poppenhaeger2022}. Bottom: same as the top panel, but for the planetary mass-loss rate.}
		\label{fig:HDsummary}
\end{figure}

For all simulations, we considered the near-ultraviolet and near-infrared stellar emission given by \citet{fossati2022} and a planetary orbital separation of 0.0344 \citep{triaud2015}. We ran simulations for a range of stellar EUV (i.e. 200--504\,\AA) and XUV (10--912\,\AA) emission flux values that encompass those derived in Section~\ref{sec:wasp80}, as well as three He/H abundance values of 0.01, 0.03, and 0.1 (by number), where the latter is the solar He/H abundance ratio. At the lower atmospheric boundary, which we locate at a pressure of 0.05\,bar, we considered a planetary atmospheric temperature of 1000\,K. To model the stellar wind, we considered the same parameters taken by \citet{fossati2022} that is a velocity of 200\,km\,s$^{-1}$, a temperature of 0.7\,MK, and a density of 10$^3$\,cm$^{-3}$ at the position of the planet, corresponding to an integral stellar mass-loss rate of 10$^{11}$\,g\,s$^{-1}$. The main difference with the work of \citet{fossati2022} lies in the significantly smaller stellar XUV emission: they employed a stellar XUV emission at 1\,AU of 7.5\,erg\,cm$^{-2}$\,s$^{-1}$ that is the largest value considered in this work. This is due to the fact that the scaling relations of \citet{poppenhaeger2022} lead to smaller EUV flux values compared to those of \citet{king2018} that were used by \citet{fossati2022}. Given the measured X-ray luminosity, the stellar wind strength expected for WASP-80 is 2$\times$10$^{13}$\,g\,s$^{-1}$ \citep{vidotto2021}. Therefore, we have further performed two additional runs considering a He/H abundance ratio of 0.1, a stellar wind mass-loss rate of 2$\times$10$^{13}$\,g\,s$^{-1}$, and the two XUV flux values (221 and 1520\,erg\,cm$^{-2}$\,s$^{-1}$) computed from the scaling relations of \citet{poppenhaeger2022} either assuming a low or high [Fe/O] coronal abundance.

The detailed results of the hydrodynamic simulations, such as the density distribution and temperature profile, resemble those presented by \citet{fossati2022}. As an example, Figure~\ref{fig:density_map} shows the proton density distribution in the planetary orbital plane obtained from the run computed considering a He/H abundance ratio of 0.01 and an EUV stellar flux at 1\,AU of 0.7\,erg\,cm$^{-2}$\,s$^{-1}$. The map shows the presence of two gas streams departing from the planet, towards (in front) and away (behind) from the planet. The planetary material after initial spherical expansion is forced to move close to the planetary orbit due to momentum conservation. The stream ahead of the planets is composed by escaped gas that feels the stellar gravitational pull and stops as a result of the interaction with the stellar wind. The stream behind the planet, also composed by planetary escaped gas, is a typical characteristic of close-in giant planets with an escaping atmosphere \citep[e.g.][]{bourrier2016,esquivel2019,ildar2018,mccann2019,debrecht2020,carolan2021,macleod2022}. Figure~\ref{fig:HeIabsorptionprofiles} shows as an example the absorption profiles obtained from time averaging \citep[see][]{dossantos2022} in the $-$0.1 to $+$0.1 planetary orbital phase range, which is the same range taken into account to extract the observed transmission spectrum from the data \citep[see Figure~1 of ][]{fossati2022}, and considering a He/H abundance ratio of 0.01. We note that averaging reduces the peak absorption by less that 5\%. Interestingly, in the case WASP-80b, we find that the cometary tail forming behind the planet does not produce significant metastable He{\sc i} absorption, in contrast to what found by \citet{macleod2022} for other systems.

Given the similarities of the detailed results with those of the simulations presented by \citet{fossati2022}, we focus here on the obtained planetary metastable He{\sc i} absorption and mass-loss rate values. All results derived from the HD simulations are summarised in Table~\ref{tab:results} and displayed in Figure~\ref{fig:HDsummary}. The full width at half maximum (FWHM) listed in Table~\ref{tab:results} is a measure of the velocity of the He{\sc i} metastable atoms along the line of sight, and thus it is directly linked to the structure and asymmetry of the absorbing atmosphere. This is key information that can be accurately extracted and compared with observations exclusively employing 3D simulations such as those adopted in this work.

We find that both metastable He{\sc i} absorption and mass-loss rate increase roughly linearly with increasing high-energy stellar emission and the He{\sc i} absorption is strongly dependent on the He/H abundance ratio. Furthermore, with increasing He/H abundance ratio, the mean molecular weight increases, which leads to a decrease of the pressure scale height, and thus of the atmospheric extension and consequently of the mass-loss rate. 

Considering a stellar EUV emission at the planetary orbit of 1520\,erg\,cm$^{-2}$\,s$^{-1}$ (i.e. the higher of the two obtained from the scaling relations; see Section~\ref{sec:wasp80}), we find that the non-detection of metastable He{\sc i} absorption implies a He/H abundance ratio smaller than ten times sub-solar. Instead, with the lower EUV stellar emission value, which is favoured by the measured \logR\ value, we obtain that metastable He{\sc i} absorption would have been undetectable for a solar He/H abundance ratio in combination with a stellar wind stronger than that expected on the basis of the measured X-ray luminosity, or for a slightly sub-solar He/H abundance ratio, or for a combination of the two. For a solar He/H abundance ratio and the foreseen stellar wind strength, the metastable He{\sc i} detectability level of the observations corresponds to a stellar XUV flux value that is about 1.5 times lower than the smaller one obtained following the scaling relations of \citet{poppenhaeger2022}. As expected, an increase in the stellar wind mass-loss rate leads to a decrease in the metastable He{\sc i} absorption signal \citep{vidotto2020,fossati2022}, however it does not appear to be enough to explain the non-detection without the need of a sub-solar He/H abundance ratio or of a stellar wind stronger than that foreseen on the basis of the measured X-ray luminosity or of both. 
\subsection{Magnetised planet}
To test the suggestion of \citet{vissapragada2022}, we further modelled the WASP-80 system considering a magnetised planet with a field strength of up to 1\,G (i.e. comparable to that of Jupiter). To properly study the impact of a planetary magnetic field on the absorption of metastable He{\sc i}, we employed a stellar spectral energy distribution with an XUV emission strong enough to ensure the production of enough metastable He{\sc i} to lead to significant absorption. Therefore, for the MHD simulations, we considered the stellar SED used by \citet{fossati2022}, namely an XUV and EUV flux equal to the strongest one considered for the HD simulations. To isolate the effect of the planetary magnetic field, we take a sufficiently low He/H abundance ratio of 0.01, so that it does not influence the formation of the planetary wind and its interaction with the stellar wind, yet providing sufficient metastable He{\sc i} absorption. Finally, we employed a stellar wind with a mass-loss rate of 10$^{11}$\,g\,s$^{-1}$.

We run simulations for a planet with a very low magnetic field of 0.01\,G, which leads to results equivalent to those of a non-magnetised planet, and two cases with a relatively strong magnetic field of 0.5 and 1.0\,G. For the strongest planetary magnetic field, we also simulated the case of a moderately strong stellar wind, that is 20 times higher density, in which the magnetised planet generates a typical magnetosphere. The main input parameters and resulting planetary mass-loss rate and metastable He{\sc i} absorption are summarised in Table~\ref{tab:results-magnetic}.
\begin{table}[]
    \renewcommand{\arraystretch}{1.5}
    \caption{MHD simulation scenarios with corresponding key modelling parameters and resulting metastable He{\sc i} absorption values.}
    \begin{tabular}{c|cc|c|c|c}
    \hline
    \hline
N & $B_{\rm p}$ & $m_{\rm p}$ & $\dot M_{\rm sw}$ & $\dot M_{\rm p}$ & A$_{\rm HeI}$ \\
 & [G] & [$m_{\rm J}$] & [10$^{11}$\,g\,s$^{-1}$] & [10$^{10}$\,g\,s$^{-1}$] & [\%] \\
    \hline
1 & 0.01 & 0.002 &  1 & 2.9 & 1.24 \\
2 &  0.5 &   0.1 &  1 & 1.5 & 0.74 \\
3 &  1.0 &   0.2 &  1 & 0.9 & 0.57 \\
4 &  1.0 &   0.2 & 20 & 0.9 & 0.95 \\
     \hline
     \end{tabular}
\tablefoot{The first column gives the number of the model run. Columns two and three list the equatorial planetary magnetic field strength in G and the magnetic moment in units of Jupiter magnetic moment. The fourth and fifth columns give the stellar wind and planetary mass-loss rates. The last column lists the total integrated metastable He{\sc i} absorption in the $\pm$10\,km\,s$^{-1}$ interval around 10\,833.2\,\AA, following time averaging from $-$0.1 to $+$0.1 in planetary orbital phase. For these simulations we considered a He/H abundance ratio of 0.01 (i.e. ten times sub-solar).}
\label{tab:results-magnetic}
\end{table}

\begin{figure}
		\centering
		\includegraphics[width=9cm]{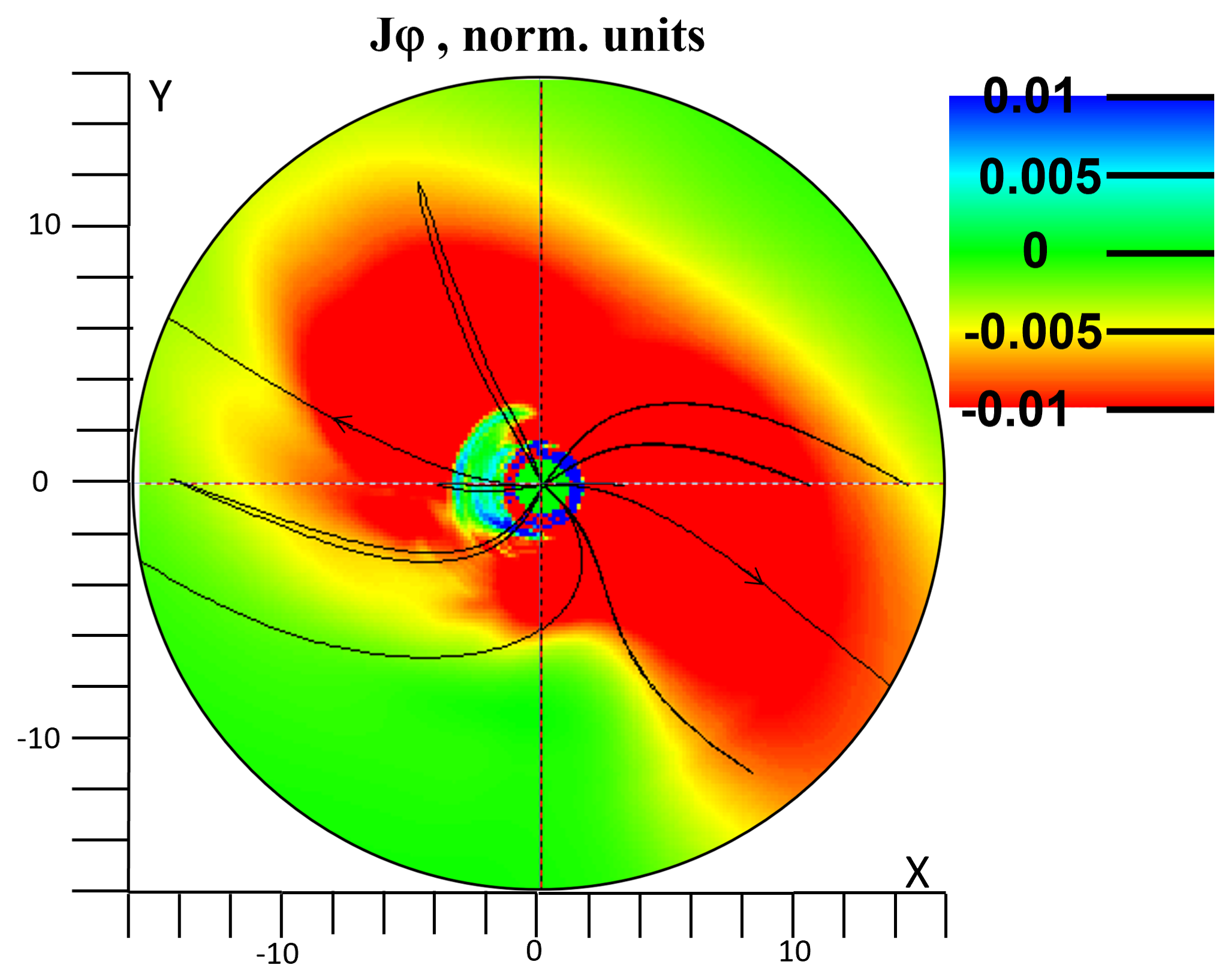}
		\includegraphics[width=9cm]{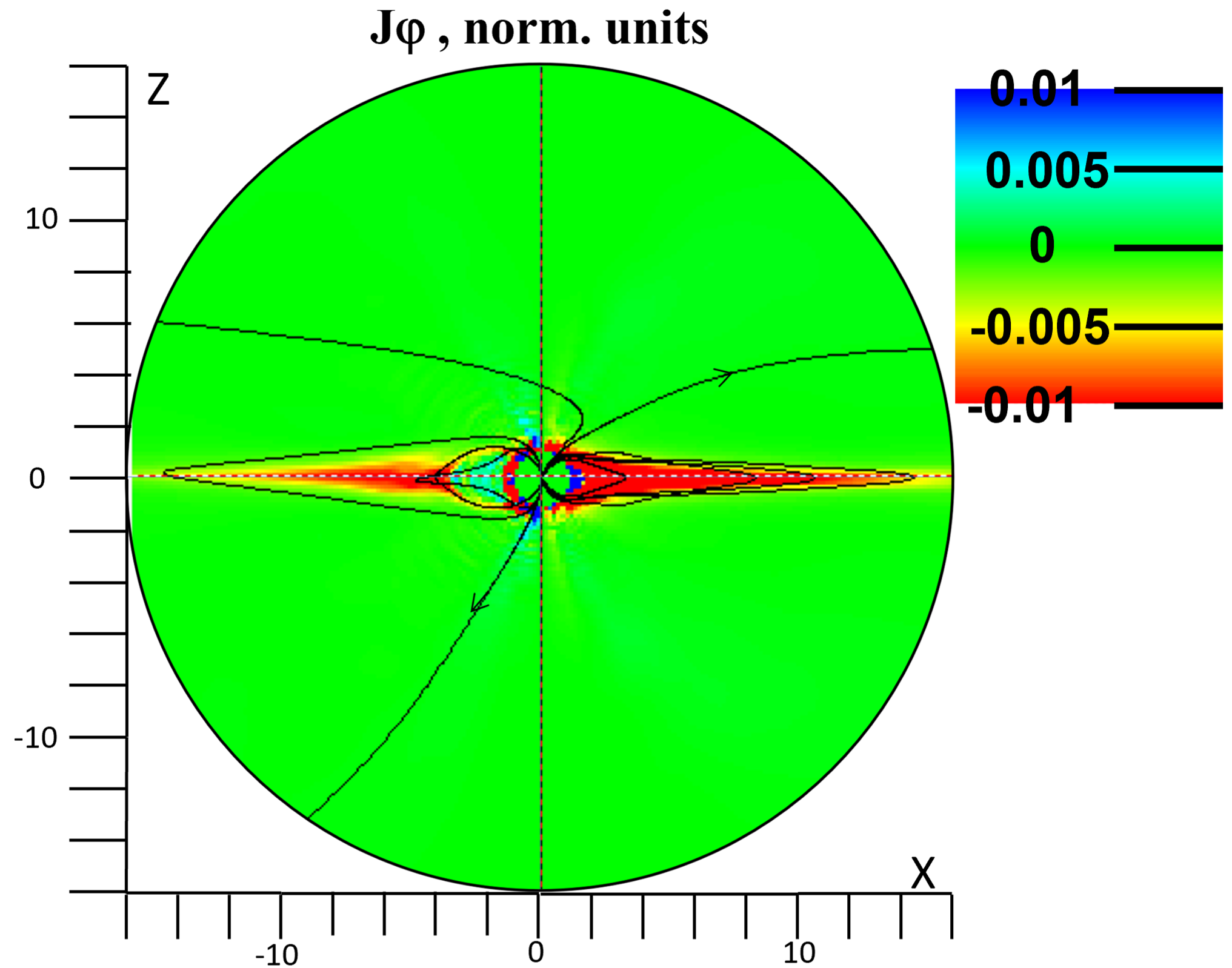}
		\caption{Distribution of the electric currents obtained from the MHD simulations. Distribution of the azimuthal component of the electric currents (J$_{\phi}$) in the orbital (top) and meridional (bottom) planes in the case of the weak stellar wind and a planetary magnetic field strength of 0.5\,G (i.e. model N2 in Table~\ref{tab:results-magnetic}). The axes are in planetary radii. The magnetic field lines are shown in black. For presentation purposes, the plots show just one eighth of the whole simulation domain. The star is located to the right at $X$\,=\,76. The current in normalised units can be converted to physical units, statampere\,cm$^{-3}$, by multiplying the values in the color bar by 0.35.}
		\label{fig:currentsB0p5}
\end{figure}

\begin{figure}
		\centering
		\includegraphics[width=9cm]{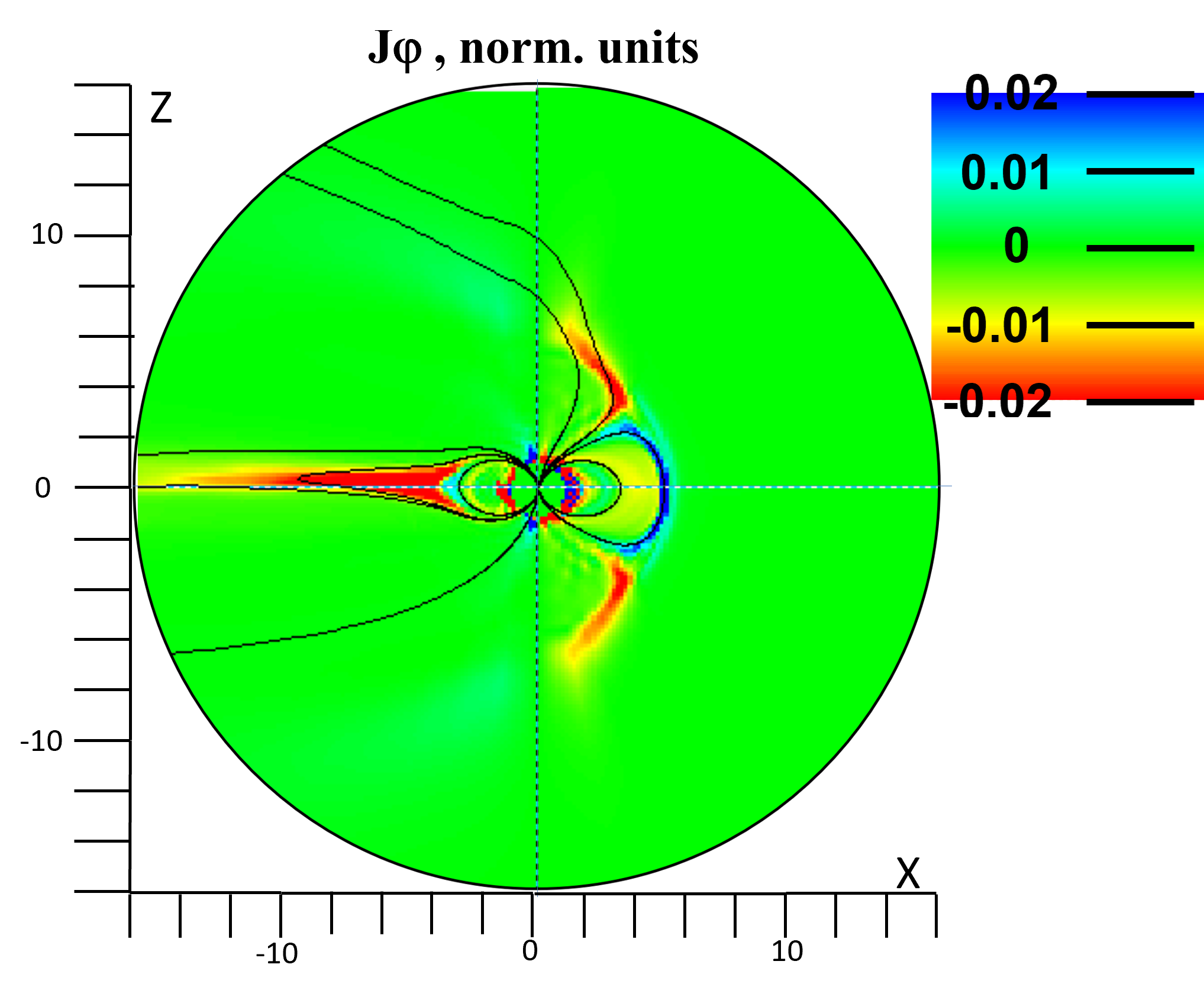}
		\caption{Same as the bottom panel of Figure~\ref{fig:currentsB0p5}, but for a planetary magnetic field strength of 1.0\,G and the strong stellar wind (i.e. model N4 in Table~\ref{tab:results-magnetic}). The star is located to the right at $X$\,=\,76.}
		\label{fig:currentsB1}
\end{figure}

Figures~\ref{fig:currentsB0p5} and \ref{fig:currentsB1} show the electric currents generated by a magnetised planet. As obtained from previous 2D and 3D simulations \citep{khodachenko2015,khodachenko2021b}, as well as earlier semi-analytical considerations \citep{khodachenko2012}, the outflowing planetary wind gas stretches and opens the magnetic dipolar field lines, forming an equatorial current layer, that is the so called magnetodisk. 

The magnetodisk is thin and surrounds the planet in the equatorial plane, but not uniformly, because of the planetary flow clock-wise rotation due to the Coriolis force. Also, there is a cavity close to the planet, the so called dead zone, where planetary material is stagnant. The generated magnetospheric system of currents is characterised by a region around the planet dominated by the dipolar magnetic field, a current sheet at the front of the magnetosphere (i.e. the magnetopause), a current sheet at high latitudes, and a current sheet forming a tail behind the planet. The strong stellar wind generates a clear bow shock around the planet, which appears to have a structure similar to that of the magnetised solar system planets (Figure~\ref{fig:currentsB1}). 

\begin{figure}
		\centering
		\includegraphics[width=9cm]{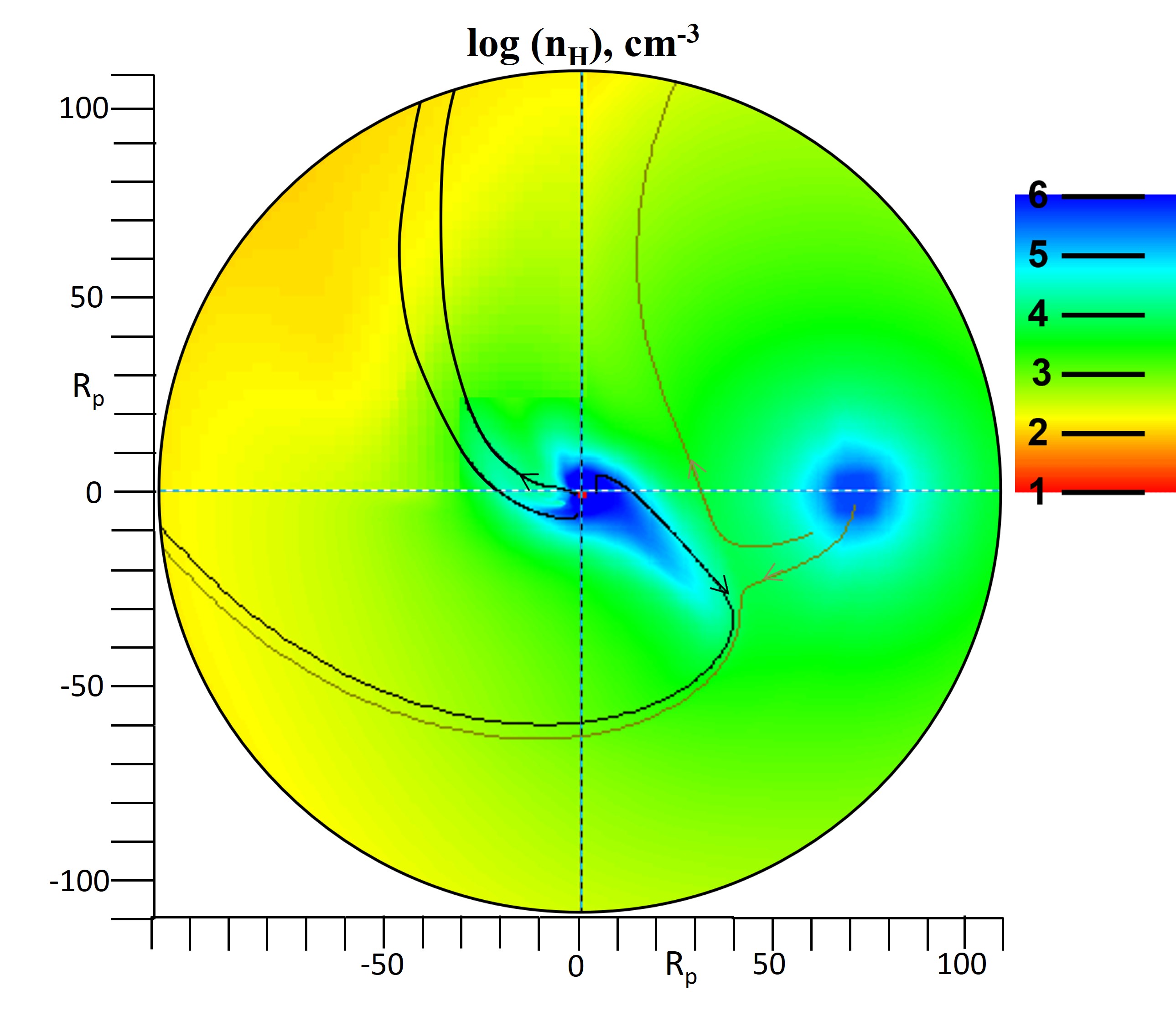}
		\caption{Proton density distribution in the orbital plane of the whole simulated domain for the run computed considering the weak stellar wind and a planetary magnetic field of 1\,G (i.e. model N3 in Table~\ref{tab:results-magnetic}). The planet is at the center of coordinate (0,0) and moves counter-clockwise relative the star, which is located at (76,0). Proton fluid streamlines originated from the planet (black) and from the star (gray) are shown. The axes are scaled in planetary radii.}
		\label{fig:density_map_Bfield}
\end{figure}

Figure~\ref{fig:density_map_Bfield} shows the proton density distribution obtained from the model run N3 (see Table~\ref{tab:results-magnetic}). A comparison between Figures~\ref{fig:density_map} and \ref{fig:density_map_Bfield} indicates that a strong planetary magnetic field truncates the tail behind the planet, which is predicted to be a prominent feature for non-magnetised planets \citep{macleod2022}. This suggests that the presence of a cometary tail could indicate the absence of a strong planetary magnetic field. The field also shortens the gas stream lying ahead of the planet further bending it towards the star.

\begin{figure}
		\centering
		\includegraphics[width=9cm]{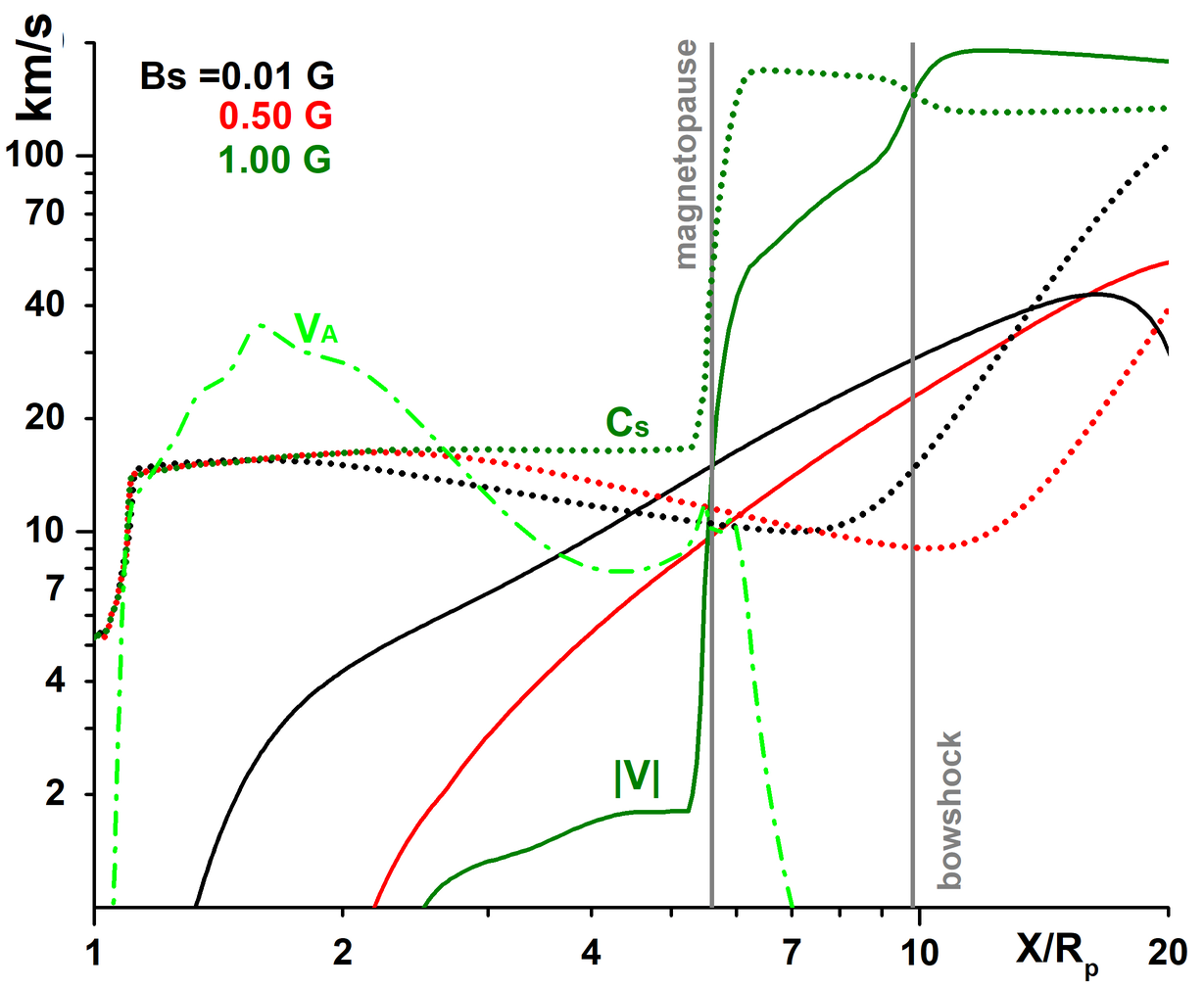}
		\includegraphics[width=9cm]{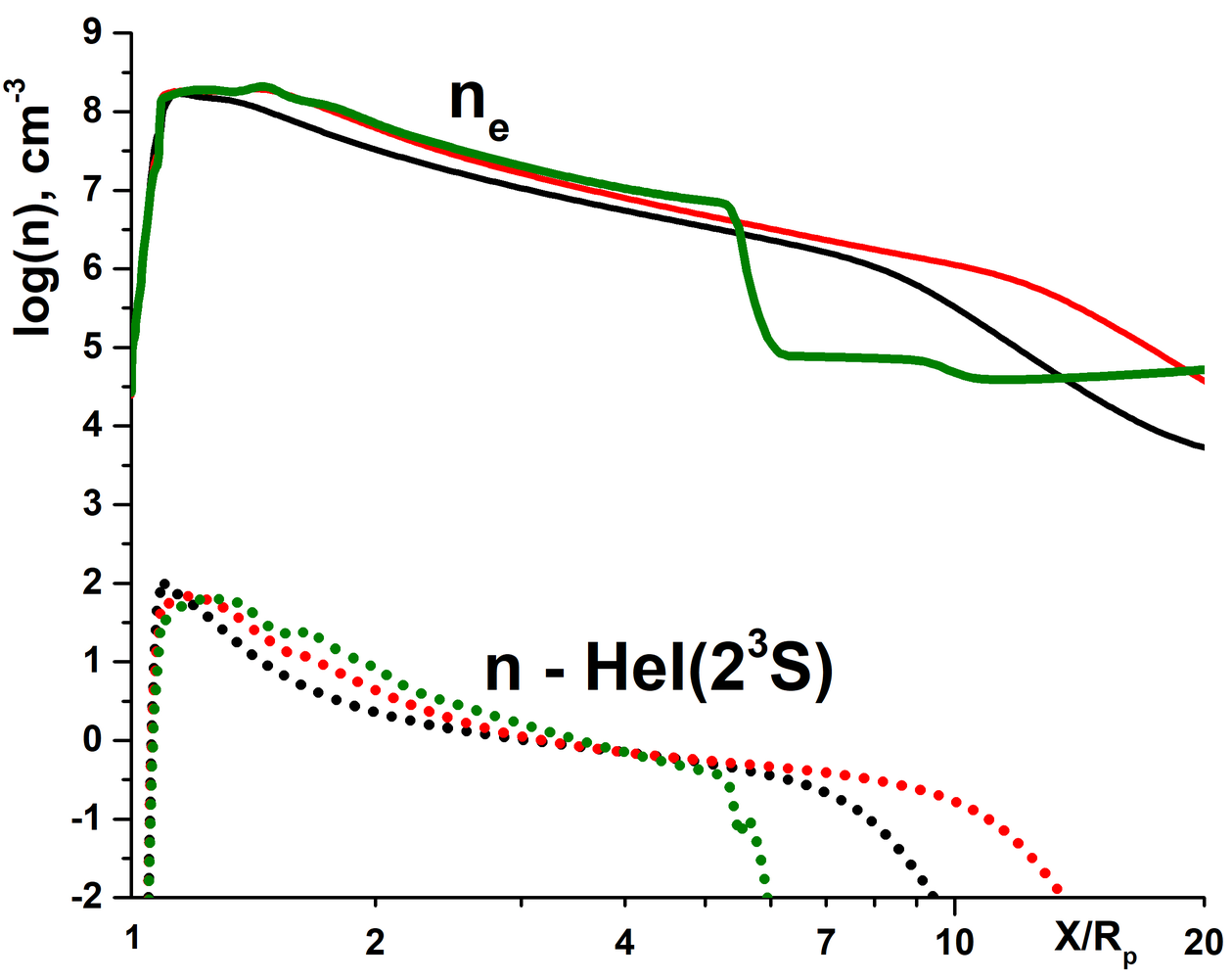}
		\caption{Profiles of the main physical quantities along the star-planet connecting line. Top: Proton bulk (solid line) and thermal (dotted line) velocity in the $X$ direction along the star-planet connecting line. The black and red lines are for the runs computed considering a planetary magnetic field of 0.01 and 0.5\,G, respectively (runs N1 and N2). The dark green lines are for the run computed considering the strongest planetary magnetic field and stellar wind (run N4). The bright green dash-dotted line shows the Alfv\'en velocity in the $X$ direction for the case of the strongest planetary magnetic field and stellar wind (run N4). Bottom: Electron (solid line) and He{\sc i} (2$^3$S) density (dotted line) profiles in the $X$ direction. As in the top panel, the black and red lines are for the runs computed considering a planetary magnetic field of 0.01 and 0.5\,G, respectively (runs N1 and N2), while the dark green lines are for the run computed considering the strongest planetary magnetic field and stellar wind (run N4).}
		\label{fig:magneticProfile}
\end{figure}

The profiles shown in Figure~\ref{fig:magneticProfile} give further details on the structure of the planetary outflow. In the case of the weaker planetary magnetic field (model run N1), the expanding planetary atmosphere is stopped relatively far from the planet, with the acceleration being driven by absorption of the stellar XUV emission and subsequent atmospheric heating. In this case, the atmosphere becomes supersonic at a distance of about 4.5 planetary radii. For a planetary magnetic field strength of 0.5\,G (model run N2), the planetary flow is strongly decelerated by the force of the magnetic tension \citep[except at the open field lines region; see ][for more details]{khodachenko2015,khodachenko2021b} and the velocity is smaller by a factor of two at distances shorter than five planetary radii compared to the case of a weakly magnetised planet. In the case of the strongest planetary magnetic field and stellar wind (model run N4), there is the formation of a magnetopause and bow shock. The position of the magnetopause, analytically computed as 
\begin{equation}
L_{\rm m}\,=\,\left(\frac{m^2}{2p_{\rm sw}}\right)^{1/6}\,\approx\,5\,R_{\rm p}\,,
\end{equation}
is close to the one obtained by the simulation (about 5.5\,$R_{\rm p}$). Within the magnetopause, the planetary wind velocity is about an order of magnitude smaller than that obtained for the case of the weakly magnetised planet. Indeed, the magnetic pressure prevails over the thermal pressure up to 3\,$R_{\rm p}$, becoming instead comparable close to the magnetopause.

\begin{figure}
		\centering
		\includegraphics[width=9cm]{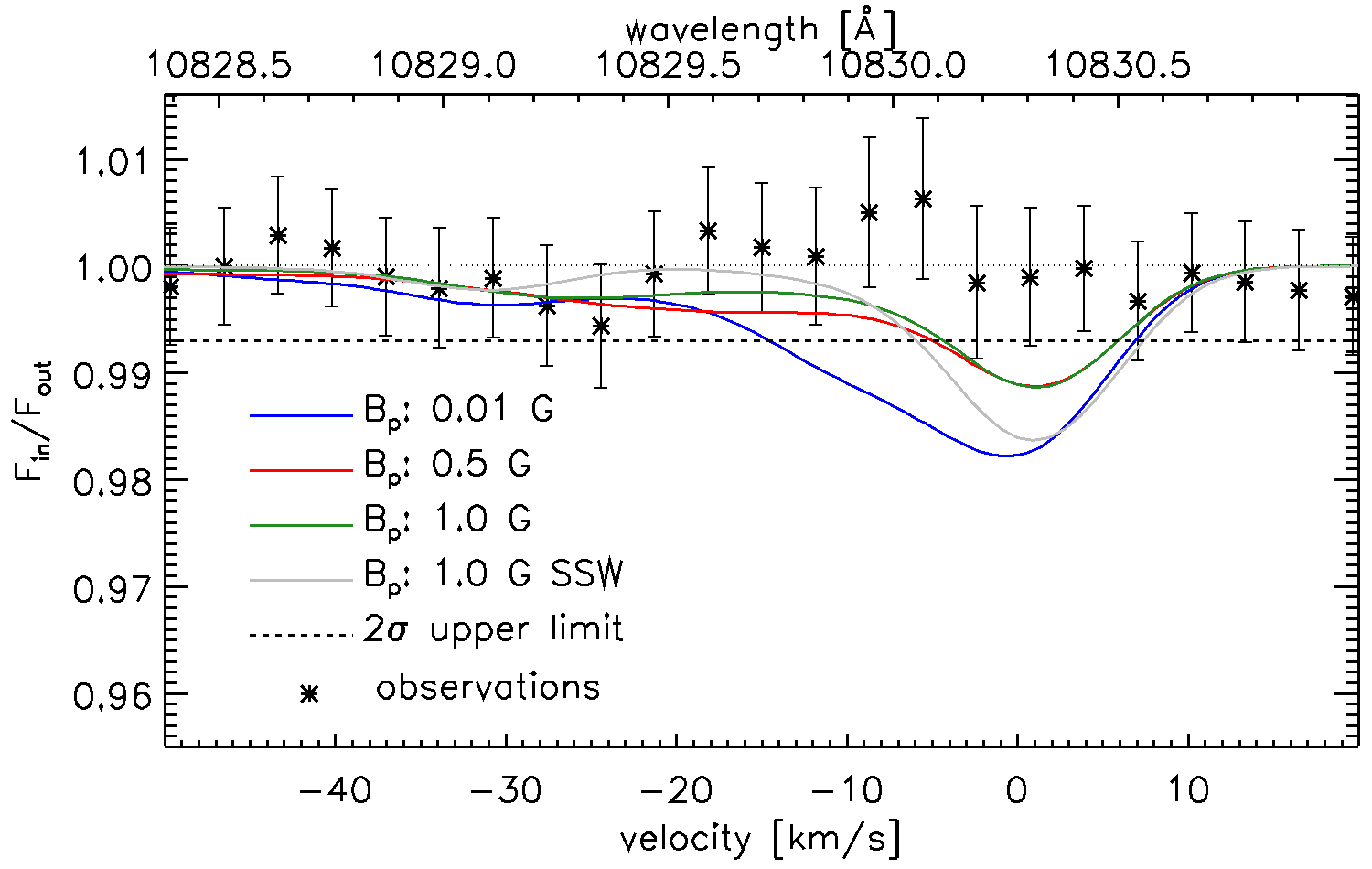}
		\caption{He{\sc i} (2$^3$S) triplet absorption profiles obtained considering different values of the planetary magnetic field at a fixed He/H abundance ratio of 0.01 in comparison with the observations (black asterisks). The blue, red, and green lines are for a stellar XUV flux at 1\,AU of 7.5\,erg\,cm$^{-2}$\,s$^{-1}$ and a planet with a magnetic field strength of 0.01, 0.5, and 1.0\,G, respectively (i.e. model runs N1, N2, N3). The gray line is for a planet with a magnetic field of 1.0\,G and a strong stellar wind (i.e. model run N4). The absorption profiles are the result of time averaging from $-$0.1 to $+$0.1 in planetary orbital phase. The zero Doppler-shifted velocity on the $x$ axis corresponds to a wavelength of 10\,830.25\,\AA. The horizontal dashed line marks the 2$\sigma$ upper limit derived from the observations \citep{fossati2022}. The horizontal dotted line at 1.00 is for reference.}
		\label{fig:magneticHeIabsorptionprofiles}
\end{figure}

Figure~\ref{fig:magneticHeIabsorptionprofiles} shows the synthetic metastable He{\sc i} absorption profiles in comparison to the observations of \citet{fossati2022}. Despite the planetary magnetic field has significantly modified the structure of the upper atmosphere, compared to the almost non-magnetised case, the He{\sc i} absorption decreases by less than a factor of two. However, with a stronger stellar wind the absorption increases, reaching almost the same level as that obtained for the almost non-magnetised planet. This occurs, because with the stronger stellar wind the magnetosphere compresses, increasing the overall density of the absorbing material around the planet.

\begin{figure}
		\centering
		\includegraphics[width=9cm]{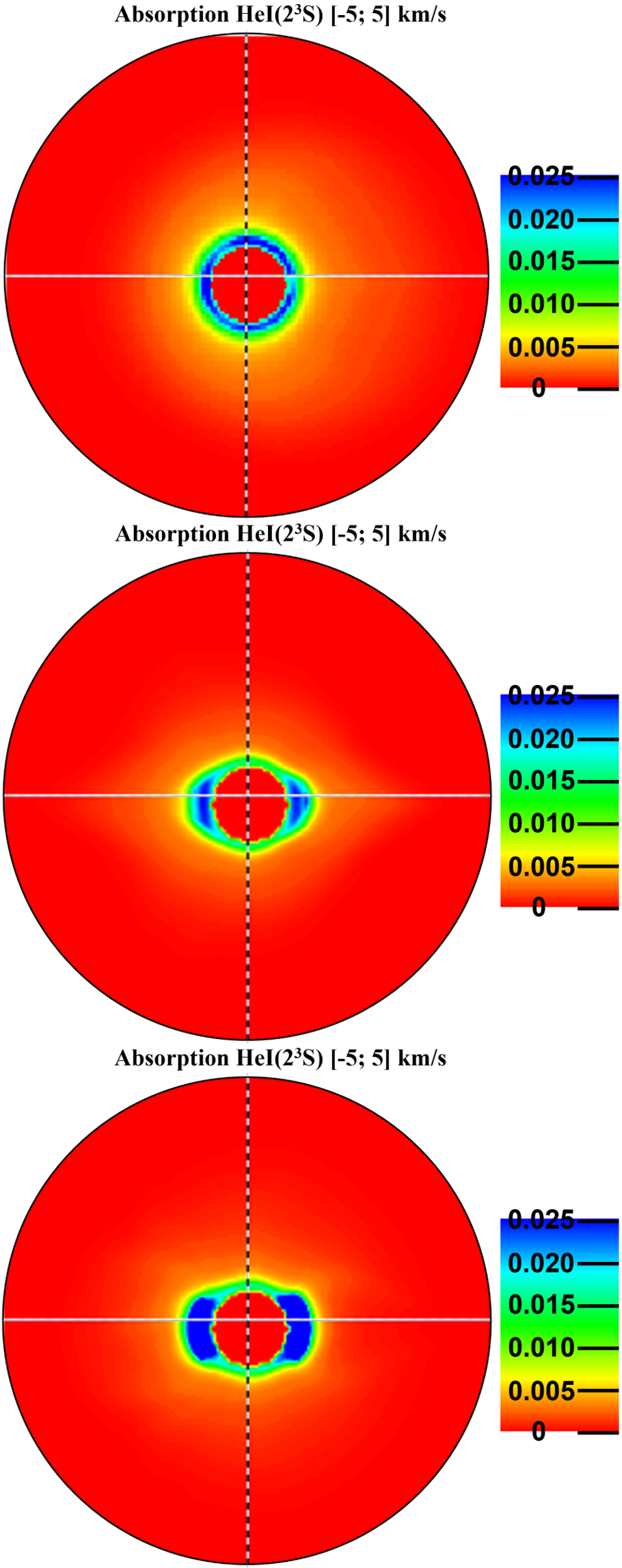}
		\caption{Distribution of the absorption of the metastable He{\sc i} line at $\approx$10830\,\AA\ across the stellar disk integrated in the $\pm$5\,km\,s$^{-1}$ range as seen by an Earth-based observer at mid-transit resulting from the N1 (top), N2 (middle), and N4 (bottom) simulations.}
		\label{fig:magnetisedHeIabsorptionmaps}
\end{figure}

These results indicate that a planetary magnetic field is unlikely to be the source of the non-detection of metastable He{\sc i} absorption in the atmosphere of WASP-80b. This is primarily, because most of the absorption (i.e. the peak absorption) takes place close to the planet for which the influence of the magnetic field is not particularly strong, while the differences in the absorption maps shown in Figure~\ref{fig:magnetisedHeIabsorptionmaps} affect mostly the line wings, which form farther away from the planet and thus do depend on the planetary magnetic field strength.

In regard of the polar regions, the simulations indicate that the planetary magnetic field reduces the outflow velocity resulting to a decreased density over the poles \citep[see][for a detailed discussion on the effect of a planetary magnetic field on the outflow velocity at the polar regions]{trammel2014,khodachenko2015,carolan2021}. However, because of the lower velocity, the planetary outflowing gas is more strongly photoionised, resulting in larger electron densities, which lead to a larger population of metastable He{\sc i} atoms produced through recombination of He{\sc ii}. Furthermore, by reducing the outflow velocity, the planetary magnetic field reduces the width of the absorption of the He{\sc i} (2$^3$S) feature from about 19\,km\,s$^{-1}$ at $B_{\rm p}$\,=\,0.01\,G to about 12\,km\,s$^{-1}$ at $B_{\rm p}$\,=\,1.0\,G. Indeed, Figure~\ref{fig:magneticProfile} shows that close to the planet the outflow velocity decreases with increasing magnetic field, but the outflow velocity is generally small and thus the reduction of the velocity does not significantly impact the absorption strength. Finally, Figure~\ref{fig:magneticHeIabsorptionprofiles} shows that with increasing planetary magnetic field from 0.5 to 1.0\,G, the population of high velocity (i.e. from about $-$30 to $-$10\,km\,s$^{-1}$) metastable He{\sc i} atoms decreases significantly, in agreement with the previous analysis.
\section{Discussion}\label{sec:discussion}
We place our results in the context of observations published for other systems. We collected the physical properties of the systems for which either measurements or non-detections of metastable He{\sc i} absorption have been published (Table~\ref{tab:systems}). For consistency, we re-analysed archival X-ray data for some of the systems and applied the scaling relations of \citet{poppenhaeger2022} to all of them. 
\subsection{High-energy emission}\label{sec:xray}
We gathered the system parameters from the literature, giving priority to more recent and/or homogeneous sources. For the systems considered by \citet[][KELT-9, HD209458, WASP-127, HD97658, HD189733, HAT-P-11, WASP-69, WASP-107, GJ9827, GJ3470, GJ436, GJ1214]{poppenhaeger2022}, we took their reported X-ray luminosity, while for the other systems we derived the X-ray luminosity from archival observations (Table~\ref{tab:xrays}). To this end, we searched for X-ray observations in the \xmm\ and \chandra\ archives. All necessary data, except for WASP-52, were found in pointed \xmm\ observations or in the slew survey. WASP-52 was observed with \chandra\ for about 10\,ks. For the targets in pointed and publicly available \xmm\ observations (WASP-80, 55\,Cnc, HAT-P-32, and Trappist-1), we reduced the datasets with SAS version 20.0, extracted the spectra, and performed the best fit analysis using XSPEC version 12.11. 

Details about the X-ray data of WASP-80 are given in Section~\ref{sec:wasp80}. In general, we used a combination of 1 or 2 APEC thermal models absorbed by a global equivalent H column ({\sc tbabs} model), where the free parameters were the temperature ($k_{\rm B}T$), the normalisation of each component, the N$_H$ equivalent column gas absorption, and the global abundances ($Z/Z_\odot$). However, for WASP-80 we kept fixed N$_H$ and $Z/Z_\odot$, as motivated in Section~\ref{sec:wasp80}. For Trappist-1, observed in six different \xmm\ visits, we extracted separately the pn and MOS spectra, combining them with SAS to obtain an average spectrum with the highest count statistics to then perform a simultaneous fit. For WASP-52 we accumulated the count rate in a region 5\arcsec\ wide and used PIMMS\footnote{{\tt https://heasarc.gsfc.nasa.gov/docs/software\\/tools/pimms.html}} to estimate its flux, using a single APEC model with $k_{\rm B}T$\,=\,0.1\,keV and solar abundance. For the undetected stars in \xmm, we computed the flux starting from the pn count rate in the 0.2--12\,keV band, using an optically thin thermal APEC model at 0.1\,keV with solar abundances and $N_{\rm H}=10^{20}$\,cm$^{-2}$ to estimate the unabsorbed flux in the 0.2--10\,keV band. For V\,1298\,Tau, we adopted the results of \citet{maggio2022}, who employed the same data analysis technique we used for the other stars.

As for WASP-80, we estimated the stellar EUV emission in the 200--504\,\AA\ wavelength range starting from the X-ray measurements and considering the scaling relations of \citet{poppenhaeger2022}. The [Fe/O] abundance of the stellar corona has been measured from the available X-ray spectra just for a few stars in our sample. For the systems in common, we took the [Fe/O] abundance value given by \citet{poppenhaeger2022}, while for the other systems we estimated the [Fe/O] abundance based on the activity and age of the host stars (Table~\ref{tab:age}). For each star in the sample, we estimated the age by using the isochrone placement algorithm briefly described in Section~\ref{sec:wasp80} \citep{bonfanti15,bonfanti16}. 

We derived the isochronal age for each star in the sample, except for Trappist-1 and GJ1214 that are extremely low-mass stars. For stars of this type, the stellar parameters are almost constant over time. For example, PARSEC models predict maximum $T_{\mathrm{eff}}$ variations of $\sim$\,0.2\% on a time scale comparable to the age of the universe. Therefore, it is not possible to infer the age from evolutionary models, however other indicators may help distinguish between young and old stars.

\citet{filippazzo15} studied a sample of ultra-cool dwarfs, also containing Trappist-1. They evaluated the ages according to youth indicators, such as the membership to nearby young moving groups \citep[see e.g.][]{gagne15} or the $\beta$-$\gamma$ gravity suffix that can be inferred from spectra \citep{kirkpatrick05,cruz09}. In the case of Trappist-1, no clear youth signatures were detected, yielding to a lower age limit of 0.5\,Gyr. For GJ1214, \citet{berta11} did not detect any kind of activity-induced chromospheric emission in either H$\alpha$ or the Na{\sc i}\,D lines. They estimated the stellar rotation period being a multiple of 53\,d, which suggests a low magnetic activity. By applying the relation from \citet{west08} between magnetic activity and kinematic age, they inferred an age greater than 3\,Gyr and confirmed this by computing the stellar motion in the $(U, V, W)$ velocity space finding $(-47, -4, -40)$\,km\,s$^{-1}$, which is consistent with membership to the Galactic old disk.
\subsection{Comparison with WASP-80b}\label{sec:comparison}
%
\begin{figure*}[h!]
		\centering
		\includegraphics[width=17cm]{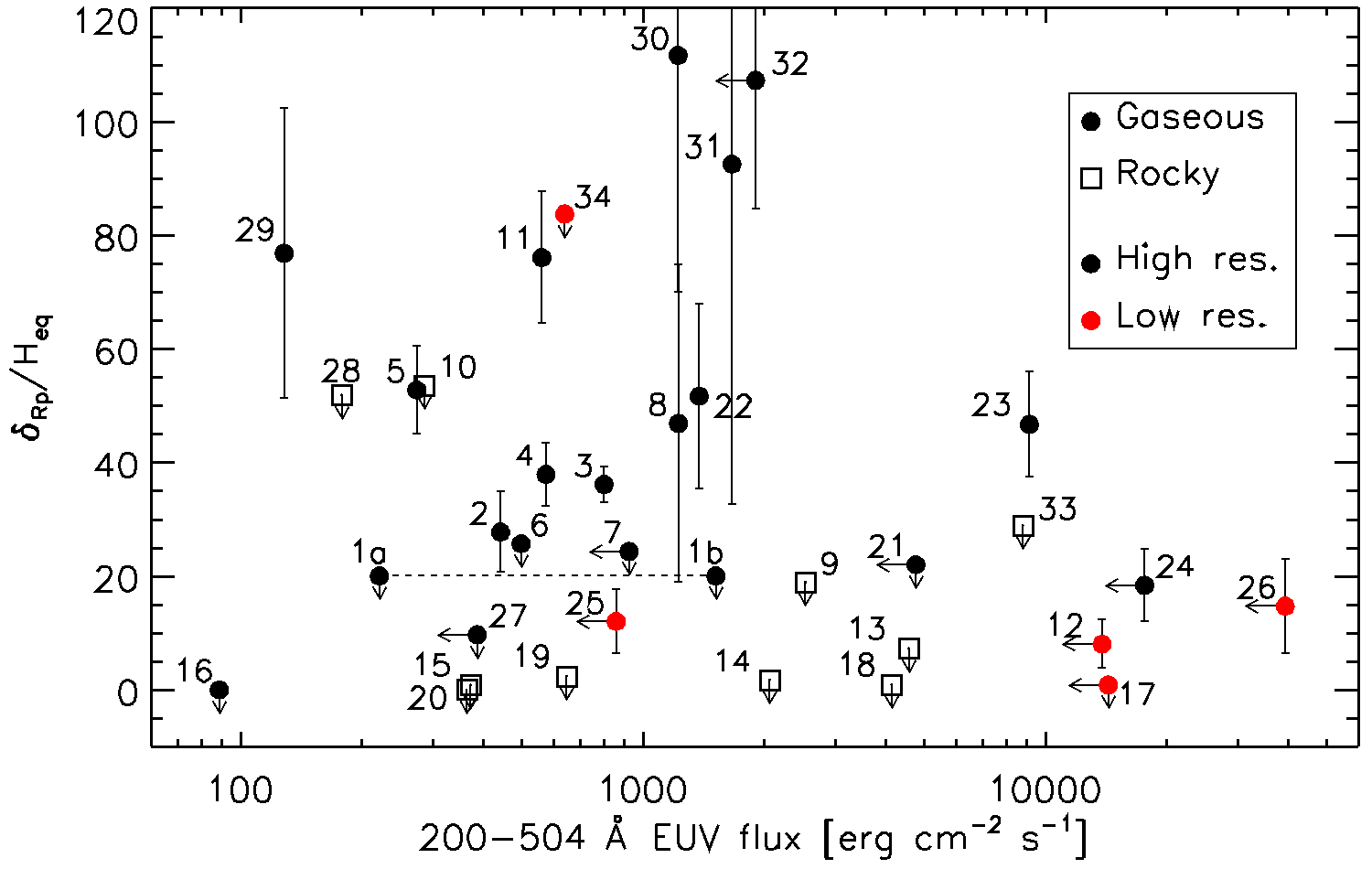}
		\caption{Size of the measured He{\sc i} absorption signal, normalised to the atmospheric scale height computed considering the planetary parameters listed in Table~\ref{tab:systems} and a mean molecular weight of a pure hydrogen atmosphere, as a function of the incident stellar EUV flux (in logarithmic scale) in the 200--504\,\AA\ wavelength range. Arrows indicate upper limits. The numbers close to each point are the labels listed in the first column of Table~\ref{tab:systems}. Gas giants are marked by filled circles, while rocky planets are marked by empty squares (we apply the term ``rocky'' to planets that presumably do not host an extensive primary hydrogen-dominated atmosphere). Black and red symbols indicate planets for which the search for metastable He{\sc i} absorption has been conducted employing high- and low-resolution techniques, respectively (low-resolution techniques comprise both low-resolution spectroscopy, $R$\,$<$\,10\,000, and narrow band photometry). The dashed horizontal line connects the two possible locations of WASP-80b, which differ solely on the assumption of a high or low [Fe/O] coronal abundance. V\,1298\,Tau\,d (\#35) is not shown, because the estimated He{\sc i} absorption signal lies significantly out of scale compared to the rest of the sample and has an uncertainty larger than 100\% (see Table~\ref{tab:systems}).}
		\label{fig:plottone}
\end{figure*}
Figure~\ref{fig:plottone} shows the size of the measured He{\sc i} absorption signal or upper limit ($\delta_{\rm Rp}$), normalised to the atmospheric scale height ($H_{\rm eq}$), as a function of incident stellar EUV flux in the 200--504\,\AA\ wavelength range \citep{poppenhaeger2022}. For each system, we estimated $H_{\rm eq}$ using the data listed in Table~\ref{tab:systems} and following \citet[][see their Section~5]{fossati2022}. The upper limits on the He{\sc i} absorption values are at the 90\% confidence level. In Figure~\ref{fig:plottone}, we divided the sample taking into account the (possible) presence of an extended hydrogen-dominated atmosphere and considering the observational technique (high- vs low-resolution). The majority of the observations reported in the literature have been collected employing ground-based high-resolution spectroscopy, except for a handful of systems that have been observed using either space-based low-resolution spectroscopy or ground-based narrow-band photometry \citep{kreidberg2018,vissapragada2022}. Also, about 70\% of the observations targeted gas giants, while the remaining observations targeted planets that probably do not host an extended hydrogen-dominated atmosphere and that for simplicity we defined ``rocky'' in Figure~\ref{fig:plottone}.

As first suggested by \citet{nortmann2018}, the systems presenting metastable He{\sc i} absorption and a measured X-ray luminosity (HD209458, HD189733, WASP-107, WASP-69, GJ3470, HAT-P-11, HAT-P-32, WASP-52, TOI560, TOI1430, TOI1683; labelled as number 2, 3, 4, 5, 8, 11, 22, 23, 29, 30, and 31, respectively) show a positive trend between the amplitude of the absorption signal and the stellar high-energy flux impinging on the planet \citep[see also][]{poppenhaeger2022}. This trend shows some scatter, particularly due to WASP-69\,b (labelled 5), HAT-P-11\,b (labelled 11), WASP-52\,b (labelled 23), TOI560\,b (labelled 29), TOI1430\,b (labelled 30), and TOI1683\,b (labelled 31) which have a He{\sc i} absorption rather different from that expected following the trend drawn by the other planets. Interestingly, TOI560\,b, TOI1430\,b, TOI1683\,b, and TOI2076\,b (the latter does not have an X-ray detection), which show a strong He{\sc i} absorption signal, are significantly less massive and smaller than the other planets in the sample, and yet likely host a primary hydrogen-dominated atmosphere. This might suggest that for sub-Neptunes the trend between metastable He{\sc i} absorption and stellar high-energy emission could be different from that drawn by Neptune and Jupiter mass planets, or that sub-Neptunes possess a high He/H abundance ratio. 

In agreement with the results of the HD simulations, in case WASP-80 has a high [Fe/O] coronal abundance ratio (labelled 1b in Figure~\ref{fig:plottone}), the non-detection of He{\sc i} is in contrast to the trend outlined by HD209458, HD189733, WASP-107, GJ3470, HAT-P-32, particularly considering that the planet is a close-in gas giant and that the host star is supposed to be of a favourable spectral type for the production of metastable He{\sc i} in the planetary atmosphere \citep{oklopcic2019}. Instead, in case WASP-80 has a low [Fe/O] coronal abundance ratio (labelled 1a in Figure~\ref{fig:plottone}), the He{\sc i} non-detection appears to be in line with the trend drawn by the other planets.
\section{Conclusions}\label{sec:conclusions}
We re-analysed archival \xmm\ observations of the planet-hosting star WASP-80. Then, we used the stellar X-ray luminosity obtained from the data as input to the scaling relations of \citet{poppenhaeger2022} to estimate the EUV flux in the 200--504\,\AA, which controls He{\sc i} metastable production, and thus absorption, in the planetary atmosphere. However, the quality of the X-ray spectrum and the large uncertainty on the stellar age did not allow us to constrain the [Fe/O] coronal abundance, which led us to consider two different EUV flux values resulting from the scaling relations, where the lower one is however favoured by the measured \logR\ value.

In light of the XUV flux values obtained for WASP-80, we run both HD and MHD simulations of the planetary upper atmosphere and of its interaction with the stellar wind to test the impact on the He{\sc i} metastable absorption signal of the XUV stellar emission, of the He/H abundance ratio in the planetary atmosphere, and of the possible presence of a planetary magnetic field. For a stellar wind about ten times weaker than solar, that is about 100 times weaker than that expected on the basis of the measured X-ray luminosity, the HD simulations revealed that He{\sc i} metastable absorption should have been detectable for a He/H abundance ratio larger than about ten times sub-solar, independently of the considered X-ray-to-EUV scaling relation. For a solar He/H abundance ratio and the lower of the two EUV flux values derived for WASP-80, which is favoured by the measured \logR\ value, the He{\sc i} non-detection can be explained by a stellar wind stronger than that expected on the basis of the measured X-ray luminosity. For the higher of the two EUV flux values derived for WASP-80, reproducing the He{\sc i} non-detection implies He/H abundance ratios smaller than ten times sub-solar. The MHD simulations indicate that the inclusion of a planetary magnetic field stronger than 0.5\,G decreases the metastable He{\sc i} absorption feature by about a factor of two, but this decrease can be compensated by a stronger stellar wind. In summary, for a low stellar [Fe/O] coronal abundance ratio (i.e. low XUV flux) and a solar He/H abundance ratio in the planetary atmosphere, the non-detection of metastable He{\sc i} absorption could be explained by the presence of a strong stellar wind. Otherwise, the non-detection would imply a sub-solar He/H abundance ratio. 

The case of WASP-80b demonstrates that the presence of an active host star, even if of favourable spectral type (i.e. K-type), cannot be used as a reliable characteristic for predicting the possible detectability of He{\sc i} in the atmosphere of a close-in giant planet. Actually, a high stellar X-ray emission might imply a too low EUV emission to produce a detectable He{\sc i} feature. As a matter of fact, the large uncertainties of scaling relations involving the inference of EUV stellar fluxes from observables \citep[X-ray, far-ultraviolet, near-ulraviolet, optical; e.g.][]{linsky2013,sreejith2020} suggest that the EUV emission of WASP-80 might simply be too small to produce a detectable He{\sc i} absorption signal, without the need to involve a sub-solar He/H abundance ratio in the planetary atmosphere or a specially strong stellar wind or the presence of a planetary magnetic field or a mix of those. The example of the WASP-80 system demonstrates the importance and urgency of bringing forward theoretical and observational studies aiming at constraining the high-energy emission of late-type stars. In particular, direct observations of the EUV spectral range, such as those proposed by the ESCAPE mission concept \citep{france2019}, would be invaluable to better understand the impact of stellar radiation on the structure and evolution of planetary atmospheres. 
\begin{acknowledgements}
Based on observations obtained with XMM-Newton, an ESA science mission with instruments and contributions directly funded by ESA Member States and NASA. This research has made use of data obtained from the Chandra Data Archive and the Chandra Source Catalog, and software provided by the Chandra X-ray Center (CXC) in the application packages CIAO and Sherpa. IFS and MSR acknowledge the support of Ministry of Science and Higher Education of the RF, grant 075-15-2020-780. GG acknowledges financial contributions from PRIN INAF 2019, and from the agreement ASI-INAF number 2018-16-HH.0 (THE StellaR PAth project). IFS and MSR acknowledge funding from ILP research project 121033100062-5 and RNF project 21-72-00129.
\end{acknowledgements}	

\bibliography{ref}	
\bibliographystyle{aa}
\begin{appendix}
\section{Properties of the considered systems with He{\sc i} absorption measurements and non-detections.}
%
\begin{sidewaystable*}
\caption{Properties of the systems for which either measurements or non-detections of the He{\sc i} metastable triplet have been published.}
\label{tab:systems}
\begin{small}
\begin{center}
\begin{tabular}{c|l|ccccc|cccc|cc}
\hline
\hline
 & Planet & $T_{\rm eff}$ & $R_{\rm s}$   & $\log{(L_{\rm X})}$ & coronal & F$_{\rm EUV}$              & $M_{\rm p}$   & $R_{\rm p}$   & $a$  & $T_{\rm eq}$ & $\left(\frac{R_{\rm p}}{R_{\rm s}}\right)^2$ & $\delta_{\rm HeI}$ \\
 &        & [K]           & [$R_{\odot}$] & [erg\,s$^{-1}$]     & iron    & [erg\,cm$^{-2}$\,s$^{-1}$] & [$M_{\rm J}$] & [$R_{\rm J}$] & [AU] & [K]          &                                              & [$R_{\rm p}$]      \\
\hline
1a & WASP-80b        & 4150$\pm$100$^1$    & 0.586$\pm$0.018$^2$    &    27.65$^3$ & l$^3$ &      221 & 0.538$\pm$0.036$^2$	 & 0.999$\pm$0.031$^2$      & 0.0344$^2$     &  816$\pm$20  & 0.03127 & $<$0.11$^4$       \\
1b & WASP-80b        & 4150$\pm$100$^1$    & 0.586$\pm$0.018$^2$    &    27.65$^3$ & h$^3$ &     1520 & 0.538$\pm$0.036$^2$	 & 0.999$\pm$0.031$^2$      & 0.0344$^2$     &  816$\pm$20  & 0.03127 & $<$0.11$^4$       \\
2  & HD209458b       & 6065$\pm$50$^1$     & 1.178$\pm$0.009$^5$    &    26.92$^6$ & l$^6$ &       35 & 0.682$\pm$0.015$^1$	 & 1.359$\pm$0.019$^1$      & 0.04707$^1$    & 1463$\pm$12  & 0.01345 &    0.29$^7$	  \\
3  & HD189733b       & 5040$\pm$50$^1$     &  0.78$\pm$0.02$^5$     &    28.30$^6$ & l$^6$ &      802 & 1.123$\pm$0.045$^1$      & 1.138$\pm$0.027$^1$      & 0.031$^1$      & 1219$\pm$13  & 0.02152 &    0.16$^{8,9}$   \\
4  & WASP-107b       & 4425$\pm$70$^{10}$  &  0.67$\pm$0.02$^{10}$  &    27.61$^6$ & h$^6$ &      575 & 0.096$\pm$0.005$^{10}$   &  0.94$\pm$0.02$^{11}$    & 0.055$^{11}$   &  757$\pm$12  & 0.01990 &    0.99$^{12,13}$ \\
5  & WASP-69b        & 4700$\pm$50$^1$     & 0.818$\pm$0.025$^5$    &    28.11$^6$ & h$^6$ &     1285 & 0.250$\pm$0.023$^1$      & 1.057$\pm$0.047$^1$      & 0.04527$^1$    &  963$\pm$11  & 0.01688 &    0.77$^{14}$    \\
6  & GJ436b          & 3479$\pm$60$^{15}$  & 0.449$\pm$0.019$^{15}$ &    26.04$^6$ & l$^6$ &       19 & 0.080$\pm$0.007$^{16}$   & 0.366$\pm$0.014$^{16}$   & 0.0308$^{16}$  &  641$\pm$11  & 0.00672 & $<$0.27$^{14}$    \\
7  & KELT9b          & 9600$\pm$400$^{17}$ & 2.418$\pm$0.058$^{17}$ & $<$27.00$^6$ & l$^6$ &    $<$78 & 2.88$\pm$0.35$^{17}$     & 1.936$\pm$0.047$^{17}$   & 0.03368$^{17}$ & 3922$\pm$165 & 0.00648 & $<$0.23$^{14}$    \\
8  & GJ3470b         & 3652$\pm$50$^{18}$  & 0.48$\pm$0.04$^{18}$   &    27.42$^6$ & l$^6$ &      147 & 0.040$\pm$0.004$^{18}$   & 0.346$\pm$0.029$^{18}$   & 0.0348$^{19}$  &  654$\pm$10  & 0.00525 &    0.96$^{20}$    \\
9  & GJ9827b         & 4340$\pm$50$^{21}$  & 0.647$\pm$0.08$^5$     &    26.81$^6$ & h$^6$ &     2536 & 0.0154$\pm$0.0015$^{21}$ & 0.1407$\pm$0.0028$^{21}$ & 0.0188$^{21}$  & 1228$\pm$26  & 0.00048 & $<$0.76$^{22}$    \\
10 & GJ9827d         & 4340$\pm$50$^{21}$  & 0.647$\pm$0.08$^5$     &    26.81$^6$ & h$^6$ &      286 & 0.0127$\pm$0.0026$^{21}$ & 0.1804$\pm$0.0041$^{21}$ & 0.0559$^{21}$  &  712$\pm$11  & 0.00079 & $<$1.94$^{22}$    \\
11 & HAT-P-11b       & 4780$\pm$50$^{23}$  & 0.769$\pm$0.048$^5$    &    27.47$^6$ & h$^6$ &      561 & 0.0736$\pm$0.0047$^{24}$ & 0.389$\pm$0.005$^{24}$   & 0.05254$^{24}$ &  882$\pm$11  & 0.00259 &    1.27$^{25,26}$ \\
12 & HAT-P-18b       & 4803$\pm$80$^1$     & 0.73$\pm$0.04$^5$      & $<$31.49$^3$ & h$^3$ & $<$13854 & 0.200$\pm$0.019$^1$	 & 0.995$\pm$0.052$^1$      & 0.05596$^1$    &  837$\pm$14  & 0.01878 &    0.17$^{27}$    \\
13 & 55 Cnc e        & 5196$\pm$24$^{28}$  & 0.95$\pm$0.08$^5$      &    27.05$^3$ & h$^3$ &     4588 & 0.0251$\pm$0.001$^{29}$  & 0.1673$\pm$0.0026$^{29}$ & 0.01544$^{29}$ & 1965$\pm$33  & 0.00031 & $<$0.34$^{30}$    \\
14 & GJ1214b         & 3250$\pm$100$^{31}$ & 0.221$\pm$0.004$^{31}$ &    25.87$^6$ & h$^6$ &     2065 & 0.0197$\pm$0.0027$^{32}$ & 0.254$\pm$0.018$^{32}$   & 0.01411$^{32}$ &  621$\pm$19  & 0.01332 & $<$0.05$^{33}$    \\
15 & HD63433b        & 5640$\pm$74$^{34}$  & 0.912$\pm$0.034$^{34}$ &    28.84$^3$ & l$^3$ &      364 & $-$  		         & 0.192$\pm$0.009$^{34}$   & 0.0719$^{34}$  &  969$\pm$13  & 0.00045 & $<$2.49$^{35}$    \\
16 & HD63433c        & 5640$\pm$74$^{34}$  & 0.912$\pm$0.034$^{34}$ &    28.84$^3$ & l$^3$ &       88 & $-$  		         & 0.2418$\pm$0.0125$^{34}$ & 0.1458$^{34}$  &  680$\pm$9   & 0.00071 & $<$1.83$^{35}$    \\
17 & WASP-12b        & 6250$\pm$100$^1$    & 1.57$\pm$0.25$^5$      & $<$29.40$^3$ & h$^3$ & $<$14353 & 1.39$\pm$0.12$^1$	 & 1.825$\pm$0.091$^1$      & 0.02312$^1$    & 2484$\pm$88  & 0.01366 & $<$0.01$^{36}$    \\
18 & Trappist-1b     & 2559$\pm$50$^{37}$  & 0.117$\pm$0.004$^{37}$ &    26.23$^3$ & h$^3$ &     4155 & 0.0032$\pm$0.0005$^{38}$ & 0.100$\pm$0.003$^{38}$   & 0.01155$^{38}$ &  393$\pm$8   & 0.00739 & $<$0.04$^{39}$    \\
19 & Trappist-1e     & 2559$\pm$50$^{37}$  & 0.117$\pm$0.004$^{37}$ &    26.23$^3$ & h$^3$ &      646 & 0.0024$\pm$0.0003$^{38}$ & 0.0812$\pm$0.0025$^{38}$ & 0.02928$^{38}$ &  247$\pm$5   & 0.00487 & $<$0.07$^{39}$    \\
20 & Trappist-1f     & 2559$\pm$50$^{37}$  & 0.117$\pm$0.004$^{37}$ &    26.23$^3$ & h$^3$ &      373 & 0.0029$\pm$0.0002$^{38}$ & 0.0933$\pm$0.0027$^{38}$ & 0.03853$^{38}$ &  215$\pm$4   & 0.00643 & $<$0.02$^{39}$    \\
21 & WASP-76b        & 6316$\pm$64$^{40}$  & 1.77$\pm$0.07$^{40}$   & $<$28.93$^3$ & l$^3$ &  $<$2011 & 0.92$\pm$0.03$^{41}$     & 1.83$\pm$0.05$^{41}$     & 0.033$^{41}$   & 2231$\pm$27  & 0.01081 & $<$0.35$^{42}$    \\
22 & HAT-P-32b       & 6269$\pm$64$^{43}$  & 1.219$\pm$0.016$^{44}$ &    28.75$^3$ & l$^3$ &     1381 & 0.585$\pm$0.031$^{44}$   & 1.789$\pm$0.025$^{44}$   & 0.0343$^{44}$  & 1816$\pm$19  & 0.02177 &    1.02$^{45}$    \\
23 & WASP-52b        & 5000$\pm$100$^{46}$ & 0.79$\pm$0.02$^{46}$   &    29.61$^3$ & l$^3$ &     9141 & 0.46$\pm$0.02$^{46}$     & 1.27$\pm$0.03$^{46}$     & 0.0272$^{46}$  & 1641$\pm$27  & 0.02613 &    0.60$^{47}$    \\
24 & WASP-177b       & 5017$\pm$70$^{48}$  & 0.885$\pm$0.046$^{48}$ & $<$30.45$^3$ & h$^3$ & $<$17647 & 0.508$\pm$0.038$^{48}$   & 1.58$\pm$0.51$^{48}$     & 0.040$^{48}$   & 1433$\pm$18  & 0.03222 &    0.26$^{47}$    \\
25 & HAT-P-26b       & 5079$\pm$88$^{49}$  & 0.788$\pm$0.071$^{49}$ & $<$27.76$^3$ & h$^3$ &   $<$858 & 0.059$\pm$0.007$^{49}$   & 0.565$\pm$0.052$^{49}$   & 0.0479$^{49}$  & 1196$\pm$17  & 0.00520 &    0.41$^{27}$    \\
26 & NGTS-5b         & 4987$\pm$41$^{50}$  & 0.739$\pm$0.013$^{50}$ & $<$30.70$^3$ & h$^3$ & $<$39491 & 0.229$\pm$0.037$^{50}$   & 1.136$\pm$0.023$^{50}$   & 0.0382$^{50}$  & 1340$\pm$9   & 0.02389 &    0.35$^{27}$    \\
27 & WASP-127b       & 5620$\pm$85$^{51}$  & 1.39$\pm$0.03$^{51}$   & $<$27.00$^6$ & l$^6$ &    $<$33 & 0.18$\pm$0.02$^{51}$     & 1.37$\pm$0.04$^{51}$     & 0.052$^{51}$   & 1575$\pm$22  & 0.00982 & $<$0.37$^{52}$    \\
28 & HD97658b        & 5212$\pm$43$^{53}$  & 0.728$\pm$0.008$^{53}$ &    27.21$^6$ & h$^6$ &      178 & 0.026$\pm$0.003$^{53}$   & 0.189$\pm$0.005$^{53}$   & 0.0805$^{54}$  &  916$\pm$6   & 0.00068 & $<$1.02$^{33}$    \\
29 & TOI560b         & 4511$\pm$110$^{55}$ & 0.65$\pm$0.02$^{55}$   &    28.00$^3$ & l$^3$ &      128 & 0.03$\pm$0.01$^{55}$     & 0.249$\pm$0.001$^{55}$   & 0.0604$^{55}$  &  999$\pm$18  & 0.00148 &    1.63$^{56}$    \\
30 & TOI1430b        & 5067$\pm$60$^{56}$  & 0.784$\pm$0.016$^{56}$ &    29.12$^3$ & h$^3$ &     1223 & 0.022$\pm$0.006$^{56}$   & 0.187$\pm$0.018$^{56}$   & 0.0705$^{56}$  & 1016$\pm$10  & 0.00058 &    2.77$^{56}$    \\
31 & TOI1683b        & 4539$\pm$100$^{56}$ & 0.636$\pm$0.024$^{56}$ &    27.87$^3$ & h$^3$ &     1665 & 0.025$\pm$0.006$^{56}$   & 0.205$\pm$0.027$^{56}$   & 0.036$^{56}$   & 1280$\pm$21  & 0.00105 &    2.50$^{56}$    \\
32 & TOI2076b        & 5200$\pm$70$^{57}$  & 0.762$\pm$0.016$^{57}$ & $<$29.68$^3$ & l$^3$ &  $<$1907 & 0.028$\pm$$-$$^{56}$     & 0.225$\pm$0.004$^{57}$   & 0.0631$^{57}$  & 1058$\pm$12  & 0.00088 &    2.69$^{56}$    \\
33 & TOI1807b        & 4730$\pm$75$^{58}$  & 0.690$\pm$0.036$^{58}$ &    28.60$^3$ & l$^3$ &     8801 & 0.0081$\pm$0.0016$^{58}$ & 0.122$\pm$0.008$^{58}$   & 0.012$^{58}$   & 2309$\pm$35  & 0.00032 &    2.70$^{59}$    \\
34 & V\,1298\,Tau\,b & 5050$\pm$100$^{60}$ & 1.278$\pm$0.070$^{60}$ & 30.23$^{61}$ & l$^3$ &      639 & 0.64$\pm$0.19$^{60}$     & 0.868$\pm$0.056$^{60}$   & 0.1719$^{60}$  &  830$\pm$13  & 0.00466 & $<$0.27$^{62}$    \\
35 & V\,1298\,Tau\,b & 5050$\pm$100$^{60}$ & 1.278$\pm$0.070$^{60}$ & 30.23$^{61}$ & l$^3$ &     1553 & $<$0.31$^{60}$           & 0.574$\pm$0.041$^{60}$   & 0.1103$^{60}$  & 1037$\pm$17  & 0.00204 &    2.40$^{62}$    \\
\hline
\end{tabular}
\tablefoot{The sixth column gives the [Fe/O] coronal abundance (i.e. h: high; l: low) considered for the computation of the stellar EUV flux incident on the planet in the 200--504\,\AA\ wavelength range and listed in column seven. Column 13 gives the effective absorption of the He{\sc i} triplet. References: 1. \citet{bonomo2017}; 2. \citet{triaud2015}; 3. This work; 4. \citet{fossati2022}; 5. \citet{gaiaedr3_2}; 6. \citet{poppenhaeger2022}; 7. \citet{alonso2019}; 8. \citet{salz2018}; 9. \citet{guilluy2020}; 10. \citet{piaulet2021}; 11. \citet{anderson2017}; 12. \citet{allart2019}; 13. \citet{kirk2020}; 14. \citet{nortmann2018}; 15. \citet{bourrier2018a}; 16. \citet{lanotte2014}; 17. \citet{borsa2019}; 18. \citet{kosiarek2019}; 19. \citet{bonfils2012}; 20. \citet{palle2020}; 21. \citet{rice2019}; 22. \citet{carleo2021}; 23. \citet{bakos2010}; 24. \citet{yee2018}; 25. \citet{allart2018}; 26. \citet{mansfield2018}; 27. \citet{vissapragada2022}; 28. \citet{vonbraun2011}; 29. \citet{bourrier2018b}; 30. \citet{zhang2021}; 31. \citet{gillon2014}; 32. \citet{harpsoe2013}; 33. \citet{kasper2020}; 34. \citet{mann2020}; 35. \citet{zhang2022a}; 36. \citet{kreidberg2018}; 37. \citet{gillon2017}; 38. \citet{grimm2018}; 39. \citet{krishnamurthy2021}; 40. \citet{tabernero2021}; 41. \citet{ehrenreich2020}; 42. \citet{casasayas2021}; 43. \citet{zhao2014}; 44. \citet{hartman2011}; 45. \citet{czesla2022}; 46. \citet{hebrard2013}; 47. \citet{kirk2022}; 48. \citet{turner2019}; 49. \citet{hartman2011}; 50. \citet{eigmuller2019}; 51. \citet{lam2017}; 52. \citet{dossantos2020}; 53. \citet{ellis2021}; 54. \citet{rosenthal2021}; 55. \citet{barragan2022}; 56. \citet{zhang2022b}; 57. \citet{osborn2022}; 58. \citet{nardiello2022}; 59. \citet{gaidos2022}; 60. \citet{suarez2022}; 61. \citet{maggio2022}; 62. \citet{vissapragada2021} }
\end{center}
\end{small}
\end{sidewaystable*}
\FloatBarrier

\section{X-ray luminosity from \xmm\ and \chandra\ observations.}
\begin{table*}[]
    \caption{Log of the X-ray observations of the sample analysed in this work.}
    \resizebox{\textwidth}{!}{
     \begin{tabular}{l|cc|ccc|cc}
     \hline
     \hline
Name & parallax & distance & Satellite & Observation & Instrument & log $F_X$ & log $L_X$ \\
     & [mas]    &     [pc] &         &               &            & [\fxu]    & [\lxu]  \\
\hline
WASP-80 	 & 	20.1141	 & 	49.7	 & 	XMM	 & 	ALL-XMMM	 & 	EPIC	 & 	-13.77	 & 	27.7	 \\
HAT-P-18	 & 	6.1863	 & 	161.6	 & 	XMM	 & 	Slew      	 & 	PN	 & 	$\leq-11$	 & 	$\leq31.49$	 \\
55 Cnc   	 & 	79.4482	 & 	12.6	 & 	XMM	 & 	0551020801	 & 	PN	 & 	$-$13.22	 & 	27.05	\\
HD63433 	 & 	44.6848	 & 	22.4	 & 	XMM	 & 	0882870101	 & 	PN	 & 	$-$11.94	 & 	28.84	\\
WASP-12 	 & 	2.4213	 & 	413.0    & 	XMM	 & 	0853380101	 & 	M2	 & 	$\leq-13.91$	 & 	$\leq29.4$	 \\
WASP-76 	 & 	5.2899	 & 	189.0    & 	XMM	 & 	0853380501	 & 	M1	 & 	$\leq-13.7$	 & 	$\leq28.93$	\\
HAT-P-32	 & 	3.4938	 & 	286.2	 & 	XMM	 & 	0853381001	 & 	PN	 & 	$-$13.24	 & 	28.75	 \\
Trappist-1	 & 	80.2123	 & 	12.5	 & 	XMM	 & 	ALL-XMM     	 & 	EPIC	 & 	$-$14.04	 & 	26.23	\\
WASP-52 	 & 	5.7262	 & 	174.6	 & 	Chandra	 & 	     15728	 & 	ACIS	 & 	$-$12.92	 & 	29.61	\\
WASP-177	 & 	5.8129	 & 	172.0	 & 	XMM	 & 	Slew	 & 	PN	 & 	$\leq-12.1$	 & 	$\leq30.45$	 \\
HAT-P-26	 & 	6.9995	 & 	142.0	 & 	XMM	 & 	0804790101	 & 	PN	 & 	$\leq-14.63$	 & 	$\leq27.76$	 \\
NGTS-5  	 & 	3.2114	 & 	311.4	 & 	XMM	 & 	Slew	 & 	PN	 & 	$\leq-12.36$	 & 	$\leq30.7$	 \\
TOI-560  	 & 	31.6569	 & 	31.6  	 & 	XMM	 & 0882870201 & EPIC  & $-$13.07        	 & 	28.00     	\\
TOI-1430 	 & 	24.2456	 & 	41.2  	 & 	XMM	 & 0882870701 & EPIC  & $-$12.18        	 & 	29.12      	\\
TOI-1683  	 & 	19.6301	 & 	50.9  	 & 	XMM	 & 0882870501 & EPIC  & $-$13.62        	 & 	27.87   	\\
TOI-2076  	 & 	23.8052	 & 	42.0  	 & 	XMM	 &  Slew      &  PN	  &	$\leq-11.64$   	 & 	$\leq29.68$	\\
TOI-1807  	 & 	23.4804  & 	42.6     & ROSAT &            & PSPC  & $-$12.73        	 & 	28.60   \\
V\,1298\,Tau &  9.2577   &  108.0    &  XMM  & 0864340301 & EPIC  & $-$11.92            &   30.23   \\
      \hline
     \end{tabular}
    }
\tablefoot{Unabsorbed X-ray fluxes ($F_X$) at Earth and luminosities ($L_X$) are in the 0.2--10\,keV band. For WASP-80 and Trappist-1 we used all available XMM observations. The stellar parallaxes and distances are from \citet{gaiaedr3_2}. For WASP-52, we used the count rate calculated in a 5\arcsec\ region and PIMMS to estimate the flux, adopting a single APEC at 0.1 keV and solar abundances absorbed by a gas equivalent column density $N_{\rm H} = 10^{20}$ cm$^{-2}$. The data relative to V\,1298\,Tau are from \citet{maggio2022} that employed the same data analysis technique we used for the other stars.}
\label{tab:xrays}
\end{table*}
\FloatBarrier

\section{Properties of the host stars.}
\begin{table*}[]
    \caption{Stellar ages derived in this work and used to infer the stellar coronal iron abundance.}
     \begin{tabular}{c|ccc|c}
     \hline
     \hline
Name & $T_{\rm eff}$ & $R_{\rm s}$   & [Fe/H] & age   \\
     & [K]           & [$R_{\odot}$] &        & [Gyr] \\
    \hline                                                                    
WASP-80      & 4150$\pm$100 & 0.586$\pm$0.018 &    0.14$\pm$0.16$^1$    &  1.6$\pm$2.3 \\
HD209458     & 6065$\pm$50  & 1.178$\pm$0.009 &    0.00$\pm$0.05$^1$    &  4.0$\pm$0.8 \\
HD189733     & 5040$\pm$50  &  0.78$\pm$0.02  &    0.03$\pm$0.08$^1$    &  7.4$\pm$2.7 \\
WASP-107     & 4425$\pm$70  &  0.67$\pm$0.02  &    0.02$\pm$0.09$^2$    &  5.1$\pm$3.2 \\
WASP-69      & 4700$\pm$50  & 0.818$\pm$0.025 &    0.15$\pm$0.08$^1$    &  7.6$\pm$4.0 \\
GJ436        & 3479$\pm$60  & 0.449$\pm$0.019 &    0.02$\pm$0.20$^3$    &  4.5$\pm$3.4 \\
KELT9        & 9600$\pm$400 & 2.418$\pm$0.058 &    0.14$\pm$0.30$^4$    &  0.34$\pm$0.14 \\
GJ3470       & 3652$\pm$50  &  0.48$\pm$0.04  &    0.20$\pm$0.10$^5$    &  1.1$\pm$1.6 \\
GJ9827       & 4340$\pm$50  & 0.647$\pm$0.08  & $-$0.26$\pm$0.09$^6$    &  4.5$\pm$4.4 \\
HAT-P-11     & 4780$\pm$50  & 0.769$\pm$0.048 &    0.31$\pm$0.05$^7$    &  9.0$\pm$2.6 \\
HAT-P-18     & 4803$\pm$80  &  0.73$\pm$0.04  &    0.10$\pm$0.08$^1$    &  5.2$\pm$2.3 \\
55 Cnc       & 5196$\pm$24  &  0.95$\pm$0.08  &    0.31$\pm$0.04$^8$    & 12.3$\pm$1.6 \\
GJ1214       & 3250$\pm$100 & 0.221$\pm$0.004 &    0.10$\pm$0.10$^9$    &      $>$3    \\
HD63433      & 5640$\pm$74  & 0.912$\pm$0.034 &    0.07$\pm$0.10$^{10}$ &  2.1$\pm$1.9 \\
WASP-12      & 6250$\pm$100 &  1.57$\pm$0.25  &    0.32$\pm$0.12$^1$    &  2.3$\pm$0.6 \\
Trappist-1   & 2559$\pm$50  & 0.117$\pm$0.004 &    0.04$\pm$0.08$^{11}$ &      $>$0.5  \\
WASP-76      & 6316$\pm$64  &  1.77$\pm$0.07  &    0.34$\pm$0.05$^{12}$ &  2.2$\pm$0.2 \\
HAT-P-32     & 6269$\pm$64  & 1.219$\pm$0.016 & $-$0.04$\pm$0.08$^{13}$ &  2.6$\pm$0.9 \\
WASP-52      & 5000$\pm$100 &  0.79$\pm$0.02  &    0.03$\pm$0.12$^{14}$ &  7.3$\pm$4.0 \\
WASP-177     & 5017$\pm$70  & 0.885$\pm$0.046 &    0.25$\pm$0.04$^{15}$ &  7.4$\pm$4.5 \\
HAT-P-26     & 5079$\pm$88  & 0.788$\pm$0.071 & $-$0.04$\pm$0.08$^{16}$ &  7.1$\pm$4.2 \\
NGTS-5       & 4987$\pm$41  & 0.739$\pm$0.013 &    0.12$\pm$0.10$^{17}$ &  2.0$\pm$1.5 \\
WASP-127     & 5620$\pm$85  &  1.39$\pm$0.03  & $-$0.18$\pm$0.06$^{18}$ & 12.5$\pm$0.4 \\
HD97658      & 5212$\pm$43  & 0.728$\pm$0.008 & $-$0.23$\pm$0.03$^{19}$ &  4.4$\pm$2.2 \\
TOI560       & 4511$\pm$110 & 0.65$\pm$0.02   &    0.00$\pm$0.09$^{20}$ & 2.1$\pm$2.0 \\
TOI1430      & 5067$\pm$60  & 0.784$\pm$0.016 & $-$0.08$\pm$0.13$^{21}$ & 9.1$\pm$2.5 \\
TOI1683      & 4539$\pm$100 & 0.636$\pm$0.024 &    0.00$\pm$0.15$^{22}$ & 1.7$\pm$1.8 \\
TOI2076      & 5200$\pm$70  & 0.762$\pm$0.016 & $-$0.07$\pm$0.13$^{21}$ & 1.6$\pm$1.7 \\
TOI1807      & 4730$\pm$75  & 0.690$\pm$0.036 & $-$0.04$\pm$0.02$^{23}$ & 5.0$\pm$2.6    \\
V\,1298\,Tau & 5050$\pm$100 & 1.278$\pm$0.070 &    0.10$\pm$0.15$^{24}$ & 0.011$\pm$0.033 \\
    \hline
    \end{tabular}
\tablefoot{The stellar effective temperature and metallicity have the same sources (already reported in Table~\ref{tab:systems} and recall here below for convenience), while the sources of the stellar radius are listed in Table~\ref{tab:systems}. The stellar ages listed in the last column have been derived employing the isochrone placement algorithm (see Section~\ref{sec:xray}) except for GJ1214 and Trappist-1 for which the lower limits come from \citet{berta11} and \citet{filippazzo15}, respectively. We remark that the [Fe/H] given here is the photospheric iron abundance relative to solar and not the iron coronal abundance. References: 1. \citet{bonomo2017}; 2. \citet{piaulet2021}; 3. \citet{lanotte2014}; 4. \citet{borsa2019}; 5. \citet{kosiarek2019};  6. \citet{rice2019}; 7. \citet{bakos2010}; 8. \citet{vonbraun2011}; 9. \citet{gillon2014}; 10. \citet{turnbull2015}; 11. \citet{gillon2017}; 12. \citet{tabernero2021}; 13. \citet{zhao2014}; 14. \citet{hebrard2013}; 15. \citet{turner2019}; 16. \citet{hartman2011}; 17. \citet{eigmuller2019}; 18. \citet{lam2017}; 19. \citet{ellis2021}; 20. \citet{barragan2022}; 21. \citet{bochanski2018}; 22. Assumption, because no [Fe/H] estimate is available in the literature; 23. \citet{nardiello2022}; 24. \citet{suarez21}. }
\label{tab:age}
\end{table*}
\FloatBarrier

\end{appendix}	
\end{document}